\documentclass[journal]{IEEEtran}

\usepackage[noadjust]{cite}

\usepackage{amsmath, amsthm, amssymb}
\usepackage{xr-hyper}
\usepackage[hidelinks]{hyperref}
\usepackage[siunitx,cuteinductors]{circuitikz}
\usepackage{tikz}
\usetikzlibrary{shapes,arrows}
\usetikzlibrary{circuits.logic.US}

\usepackage{soul,color}
\soulregister\cite7
\soulregister\ref7
\soulregister\pageref7

\newcommand{\subparagraph}{}

\usepackage[compact]{titlesec}
\titlespacing{\section}{0pt}{*4}{*2}
\titlespacing{\subsection}{0pt}{*1}{*0.5}

\tikzstyle{block} = [draw, shape=rectangle, minimum height=3em, minimum width=3em, node distance=2cm, line width=1pt]
\tikzstyle{sum} = [draw, shape=circle, node distance=1.5cm, line width=1pt, minimum width=1.25em]
\tikzstyle{branch}=[fill,shape=circle,minimum size=4pt,inner sep=0pt]
\tikzstyle{io} = [node distance=1.4cm, minimum width=1.25em]
\tikzstyle{mix} = [shape=circle, node distance=1.5cm, line width=1pt, minimum width=24pt]

\newcommand{\Dp}{\partial}

\newcommand{\D}{\displaystyle}

\newcommand{\V}{\overline}

\renewcommand{\lim}[2]{\lim_{#1 \rightarrow #2}}

\newcommand{\C}[1]{\left[#1\right]}

\newcommand{\LL}[1]{\left\{#1\right\}}
\newcommand{\Pa}[1]{\left(#1\right)}

\ifCLASSOPTIONcompsoc
  \usepackage[caption=false,font=normalsize,labelfont=sf,textfont=sf,subrefformat=parens,labelformat=parens]{subfig}
\else
  \usepackage[caption=false,font=footnotesize,subrefformat=parens,labelformat=parens]{subfig}
\fi

\begin{document}
\begin{titlepage}
{\bfseries\LARGE IEEE Copyright Notice\par}
\vspace{1cm}
{\scshape\Large © 2019 IEEE.  Personal use of this material is permitted.  Permission from IEEE must be obtained for all other uses, in any current or future media, including reprinting/republishing this material for advertising or promotional purposes, creating new collective works, for resale or redistribution to servers or lists, or reuse of any copyrighted component of this work in other works. \par}
\vspace{1cm}
{\Large This is the author's version of an article that has been published in this journal. Changes were made to this version by the publisher prior to publication. \par}
\vspace{1cm}
{\Large The final version of record is available at http://dx.doi.org/10.1109/TMTT.2019.2912366 \par}
\end{titlepage}

\bibliographystyle{IEEEtran}
\raggedbottom

\setlength{\abovedisplayskip}{6pt}
\setlength{\belowdisplayskip}{6pt}


\title{Piecewise Semi-Analytical Formulation for the Analysis of Coupled-Oscillator Systems}

\author{Pedro~Umpierrez,~\IEEEmembership{Member,~IEEE,}
        Victor~Ara\~na,~\IEEEmembership{Member,~IEEE,}
        and~Sergio~Sancho,~\IEEEmembership{Member,~IEEE}

\thanks{This work was supported by the Spanish Ministry of Economy and Competitiveness and the European Regional Development  Fund  (ERDF/FEDER)  under Project TEC2014-60283-C3-(1/2)-R and Project  TEC2017-88242-C3-(1/2)-R}

\thanks{P. Umpierrez and V. Ara\~na are with the Department
of Signals and Communications, University of Las Palmas, Las Palmas de Gran Canaria,
35017, Spain e-mail: victor.arana@ulpgc.es.}

\thanks{S. Sancho is with the Department
of Communications Engineering, University of Cantabria, Santander,
39005, Spain e-mail: sanchosm@unican.es.}}

\markboth{IEEE TRANSACTIONS ON MICROWAVE THEORY AND TECHNIQUES}{}

\maketitle

\begin{abstract}
A new simulation technique to obtain the synchronized steady-state solutions existing in coupled oscillator systems is presented. The technique departs from a semi-analytical formulation presented in previous works. It extends the model of the admittance function describing each individual oscillator to a piecewise linear one. This provides a global formulation of the coupled system, considering the whole characteristic of each voltage-controlled oscillator (VCO) in the array. In comparison with the previous local formulation, the new formulation significantly improves the accuracy in the prediction of the system synchronization ranges. The technique has been tested by comparison with computationally demanding circuit-level Harmonic Balance simulations in an array of Van der Pol-type oscillators and then applied to a coupled system of FET based oscillators at 5 GHz, with very good agreement with measurements.
\end{abstract}


\section{Introduction}
\IEEEPARstart{C}{oupled} { oscillator systems can be used in several applications like power combination, beam-steering in phased-arrays \cite{Lynch:CO,Heath:beam,pogorzelski:continuum,moussounda:dissertation,pogorzelski:book:coupled,kijsanayotin:coupled,qi_phased_array}
or in sensor applications \cite{simeone:wireless,ponton:wireless:ims,ponton:wireless:mtt}, where the synchronization between the oscillators provides a common time-scale between the nodes of the sensor network.} {In previous works \cite{suarez:OA:noise,suarez:OA:general} a semi-analytical formulation (SAF) has been proposed to analyze the behavior of coupled oscillator systems. The SAF is a frequency-domain compact model of a voltage-controlled oscillator (VCO) under free-running conditions or in the presence of an injection source. It is based on the fact that, applying {the implicit function theorem \cite{suarez:OA:noise,suarez:OA:general}}, the whole Harmonic Balance system associated with the VCO can be compacted into a single equation in terms of a nonlinear admittance function $Y(V,\omega,\eta,G^r,G^i)$. This function is nonlinear with respect to all the variables $(V,\omega,\eta,G^r,G^i)$ which are, respectively, the first harmonic amplitude, the free-running frequency, the tuning voltage, and the real and imaginary parts of the phasor associated with the injection generator. In the SAF used so far, the function $Y$ has been approached by its first-order Taylor series about the free-running state $(V_o(\eta_0),\omega_o(\eta_0))$, corresponding to a given tuning voltage value $\eta_0$  \cite{suarez:OA:noise,suarez:OA:general,suarez:OA:global,suarez:libro2,ramirez:divider,sancho:cycle_slips}, . The admittance derivatives $Y_x$ used in this Taylor series are calculated through finite differences in circuit-level HB simulations with the aid of an auxiliary generator, as explained in \cite{suarez:libro2,ramirez:divider}. The resulting SAF can be used to model the steady state behavior of the VCO for tuning voltage values in the neighborhood of $\eta_0$. The SAF has been proven to be a powerful technique to model the VCO behavior when operating under free-running conditions \cite{suarez:libro2}, synchronized to an external source \cite{suarez:libro2,suarez:OA:general,ramirez:divider,sancho:cycle_slips} or as a part of an array of coupled oscillators, operating both in phased array antenna systems  \cite{suarez:OA:general,suarez:OA:global,suarez:OA:noise} and in wireless sensor networks \cite{simeone:wireless,ponton:wireless:ims,ponton:wireless:mtt}.} In the case of coupled oscillators, the SAF is able to predict the coexisting solutions of the array, together with their bifurcation loci. Indeed, in some array circuits the SAF can be considered as an alternative to circuit-level HB where the latter is computationally demanding and fails to converge in some cases \cite{suarez:OA:global,ponton:modes}.

{ The SAF is a general-purpose formulation that models the coupled system behavior. Therefore, any further development of this technique will be of great interest, since it will improve the simulation results of these systems in all their applications. In the present work, the SAF has been extended to simulate the case of a coupled system whose VCOs operate under non-linearizable variations of the tuning voltage. It must be noted that this scenario can take place in any application where the VCO tuning voltage needs to be swept in a given range of values, like phased-array antenna systems or, for example, chirped oscillators used to detect input-signal frequencies \cite{ramirez:chirped:tmtt}. Since the SAF approaches the admittance function $Y$ by its first-order Taylor series, this formulation lacks accuracy as any of the VCOs in the system is detuned farther from its free-running state. To overcome this limitation, a piecewise technique modeling the nonlinear variation of the admittance function $Y$ with the tuning voltage has been developed. The new technique provides a piecewise (PW) SAF for the VCO that makes use of a set of admittance derivatives that depend on the VCO tuning-voltage. This new technique has the advantage that these derivatives can be easily obtained in circuit-level HB simulations of the single VCO response to the tuning voltage. The new formulation takes into account the whole VCO characteristic, performing a global analysis, instead of the previous non-piecewise SAF that is based on a local analysis resulting from linearization about a single point of the characteristic. In addition, the technique has been tested by analyzing the asymmetric behavior of the system in the case of non-identical oscillators, which had not been considered in previous works \cite{suarez:OA:noise,suarez:OA:general}.

The paper is organized as follows. In Section II, the new piecewise (PW) SAF is detailed and compared with the previous non-PW SAF. In Section III, the technique is tested by comparison with circuit-level HB simulations in a free-running array of Van der Pol-type oscillators. Then, it is applied to a coupled system of FET based coupled oscillators at 5 GHz, both under free-running conditions and injection-locked to an external generator {\cite{adler:locking}}. This system is intended to operate in a phased array antenna system.} In beam scanning applications, the steering angle is determining by imposing a constant phase shift $\Delta\varphi$ between adjacent oscillators. The free-running frequency of each individual VCO is controlled by a tuning voltage $\eta$. As explained in \cite{Lynch:CO,Heath:beam}, the constant phase shift $\Delta\varphi$ can be continuously varied by detuning the outermost oscillators only. Since the HB technique is unable to converge in this system, in this case the results are verified through comparison with measurements.
  
\section{Modeling the oscillator array using piecewise linear admittance functions} \label{s5}

\subsection{Piecewise model of the individual VCO} \label{s3}
\begin{figure} [!h]
    \centering
\includegraphics[width=3in]{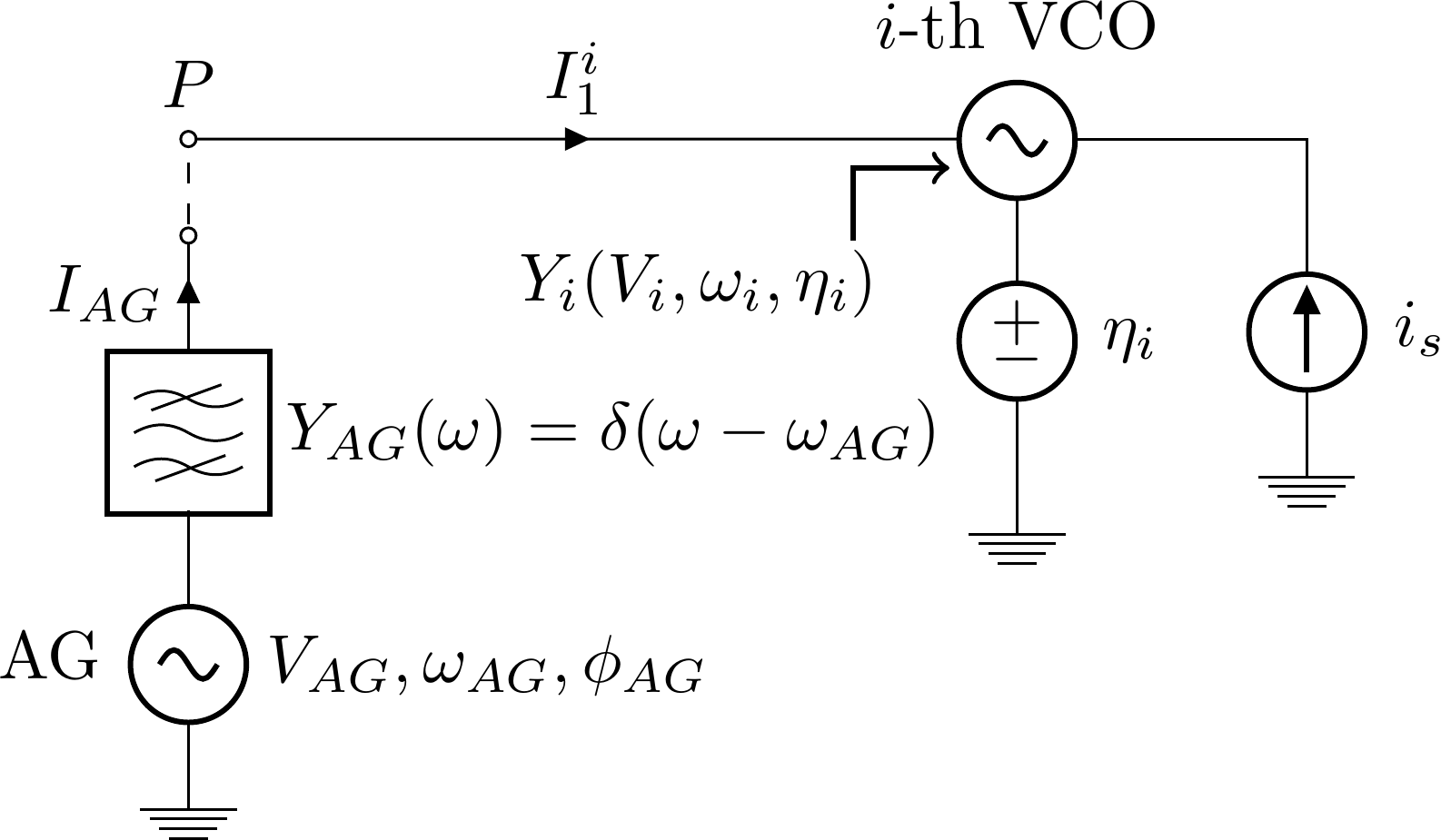}
 \caption{Procedure to extract the model of the individual $i-$th VCO. An auxiliary generator (AG) is connected to the output node $P$ to calculate the first-harmonic input admittance function $Y_i(V_i,\omega_i,\eta_i)$ and the derivatives $I_{iG_{\pm 1}}(V_i,\omega_i,\eta_i)$ providing the dependence of the input current $I_1^i$ on the injection source $i_s(t)$.}
  \label{f8}
\end{figure}

The new technique presented here to model the behavior of each individual $i-$th VCO in the array departs from the semi-analytical formulation presented in previous works \cite{suarez:OA:noise,suarez:OA:general,suarez:libro2}. The procedure is schematized in Fig. \ref{f8}. In this figure, the $i-th$ VCO of the system of coupled oscillators is represented together with a tuning voltage source $\eta_i$ and a current source $i_s(t)=I_s\cos(\omega_st+\theta_s)$ modeling the effect of an external injection source. For the sake of generality, the formulation will be derived for the general case where the injection source may be placed at any circuit node. The VCO is modeled by the first-harmonic component $I_1^i$ of the current entering the VCO from the output node $P$. This component of each VCO will be introduced later in the system of equations modeling the array behavior. 

An expression for $I_1^i$ in terms of the VCO variables can be derived in the following way. As explained in \cite{suarez:OA:noise,suarez:OA:general,suarez:libro2}, the Harmonic Balance (HB) system associated with the VCO circuit can be combined with the Implicit Function (IF) theorem to compact all the harmonic content information into a single equation. This is the equation corresponding to the the first harmonic component of the total current entering the node $P$ which, according to the Kirchhoff's current law (KCL), fulfills $I_1^i(V_i,\phi_i,\omega_i,\eta_i,G^r,G^i)=0$, where $(V_i,\phi_i,\omega_i)$ are the first harmonic amplitude, phase and frequency of the voltage signal at $P$, and $G^r,G^i$ are the real and imaginary parts of the injection current source phasor $G_1\equiv I_se^{j\theta_s}$. Note that, as described in \cite{suarez:OA:noise,suarez:OA:general,suarez:libro2,ramirez:sincro_noise}, the application of the IF theorem assures that this equation contains all the harmonic information of the VCO, reducing the number of harmonic variables to the pair $\Pa{V_i,\phi_i}$ at the cost of increasing the complexity of the function $I_1^i$. Following \cite{ramirez:sincro_noise,sancho:cycle_slips}, assuming that the injection source amplitude is small enough, the current function $I_1^i$ can be approached by a first-order Taylor series about the non-injected state as:
\
\begin{IEEEeqnarray}{lll}
	I_1^i(V_i,\phi_i,\omega_i,\eta_i,G^r,G^i)\simeq Y_i(V_i,\omega_i,\eta_i)V_ie^{j\phi_i}+\nonumber\\
	+I_s\Pa{I_{iG_1}e^{j\theta_s}+I_{iG_{-1}}e^{-j(\theta_s-2\phi_i)}},\ 
	G_{\pm 1}\equiv I_se^{\pm j\theta_s}\label{e3}\\
	I_{iG_{\pm 1}}\equiv\D\frac{\Dp I_1^i(V_i,\phi_i=0,\omega_i,\eta_i,G^r=0,G^i=0)}{\Dp G_{\pm 1}}\nonumber
\end{IEEEeqnarray}

where the current entering the VCO in the absence of injection source has been expressed in terms of the input admittance function $Y_i(V_i,\omega_i,\eta_i)$ and the influence of the external generator is determined by the functions $I_{iG_{\pm 1}}(V_i,\omega_i,\eta_i)$. In order to calculate these functions, an auxiliary generator (AG) is connected to the node $P$. This is a single-tone voltage source of amplitude, phase and frequency $(V_{AG},\phi_{AG},\omega_{AG})$. The AG is meant to affect the VCO only at the frequency component $\omega_{AG}$, which is achieved by an ideal band-pass filter of admittance $Y_{AG}(\omega)=\delta(\omega-\omega_{AG})$. 

The AG is used to determine the current $I_1^i$. For this purpose, the current source frequency is set to $\omega_s=\omega_{AG}$, modifying the first harmonic KCL equation at node $P$ as:
\
\begin{equation} \label{e7}
\begin{aligned}
	&I_1^i=
	I_{AG}(V_{AG},\phi_{AG},\omega_{AG},\eta,G^r,G^i)
\end{aligned}
\end{equation}

Now, the functions $Y$ and $I_{G_{\pm 1}}$ can be evaluated at a given point $(V,\omega,\eta)$ through circuit-level HB simulations setting the AG amplitude and frequency to $V_{AG}=V_i$ and $\omega_{AG}=\omega_i$ . In a first stage, the injection source is removed $(I_s=0)$ and the current crossing through the AG is obtained to evaluate the admittance function as $Y_i(V_i,\omega_i,\eta_i)=I_{AG}(V_i,\phi_{AG},\omega_i,\eta_i)/V_ie^{j\phi_{AG}}$. Note that, due to the oscillator autonomy, this result is independent of the AG phase $\phi_{AG}$. In a second stage, the derivatives $I_{iG_{\pm 1}}$ are obtained: The AG phase is set to $\phi_{AG}=0$ according to (\ref{e3}) and the derivatives are calculated through finite differences in $(G^r,G^i)$ using the chain rule relations:
\
\begin{IEEEeqnarray}{lll}
	I_{iG_{1}}=
	\frac{1}{2}\Pa{I_{iG^{r}}-jI_{iG^{i}}}, \
	I_{iG_{-1}}=
	\frac{1}{2}\Pa{I_{iG^{r}}+jI_{iG^{i}}},\nonumber\\
	I_{iG^{r,i}}(V_i,\omega_i,\eta_i)= \\
	 \qquad\qquad=\D\frac{\Dp I_{AG}(V_i,\phi_i=0,\omega_i,\eta_i,G^r=0,G^i=0)}
	{\Dp G^{r,i}} \nonumber
\end{IEEEeqnarray}

Under free-running conditions $(I_s=0)$, the tuning voltage $\eta_i$ determines the VCO steady state amplitude and frequency values $\Pa{V_i^o(\eta_i),\omega_i^o(\eta_i)}$. These values are a solution for the free-running VCO HB system fulfilling $I_1^i\Pa{V_i^o(\eta_i),\phi_i,\omega_i^o(\eta_i),\eta_i,G^r=0,G^i=0}=0$ and therefore $Y_i\Pa{V_i^o(\eta_i),\omega_i^o(\eta_i),\eta_i}=0$, showing that the input admittance function vanishes when evaluated at the free-running solution. The model is intended for a VCO operating in an array under weak-coupling and weak injection conditions. Then, for each value $\eta_i$ of the tuning voltage, the steady state amplitude and frequency are expected to be close to the free-running values $\Pa{V_i^o(\eta_i),\omega_i^o(\eta_i)}$. Taking this into consideration, in previous works \cite{suarez:OA:noise,suarez:OA:general,suarez:libro2} the functions $Y_i$ and $I_{iG^{r,i}}$ have been linearized about a given free-running solution $\Pa{V_i^o(\eta^c),\omega_i^o(\eta^c),\eta^c}$ corresponding to a tuning value $\eta^c$ belonging to the tuning range required for the array operation:
\
\begin{IEEEeqnarray}{lll}
	Y_i\Pa{V_i,\omega_i,\eta_i}\simeq
	Y_{iV}\Delta V_i+Y_{i\omega}\Delta\omega_i+
	Y_{i\eta}(\eta_i-\eta^c),\nonumber\\
	\Delta V_i=V_i-V_i^{o}(\eta^c),\qquad\Delta\omega_i=\omega_i-\omega_i^{o}(\eta^c),\label{e8}\\
	I_{iG^{r,i}}(V_i,\omega_i,\eta_i)\simeq
	I_{iG^{r,i}}(V_i^o(\eta^c),\omega_i^o(\eta^c),\eta^c)\equiv I_{iG^{r,i}}^c,\nonumber\\
	Y_{ix}\equiv\D\frac{\Dp Y_i\Pa{V_i^{o}(\eta^c),\omega_i^{o}(\eta^c),\eta^c}}{\Dp x},\quad x\equiv V_i,\omega_i,\eta_i\nonumber
\end{IEEEeqnarray}

where the derivatives $Y_{ix}$ are calculated through finite differences in $x$. In (\ref{e8}), the functions $Y_i,I_{iG^{r,i}}$ are approached by the hyperplanes in the three-dimensional space $(V_i,\omega_i,\eta_i)$ resulting from their first-order Taylor expansion about the free-running solution $\Pa{V_i^o(\eta^c),\omega_i^o(\eta^c),\eta^c}$. Since the coefficients of this series are the derivatives $Y_{ix}$ calculated in circuit-level HB, formulation (\ref{e8}) is called semi-analytical (SAF). 

{ Note that the SAF formulation (\ref{e8}) has been developed following the notation used in the previous papers \cite{suarez:OA:noise,suarez:OA:general,suarez:OA:global,suarez:libro2,ramirez:divider,sancho:cycle_slips}. In this formulation, the parameters $\eta_i,\ i=1,\ldots,N$ representing the dc tuning voltage of each $i-$th oscillator are the only variables associated with the dc sources. The rest of the variables are associated with the first-harmonic components at the observation node of each $i-$th oscillator.}

As the coupled system is intended to operate along a relatively wide range of tuning voltage values, approach (\ref{e8}) results inaccurate for high values of the $\eta_i-\eta^c$ increment. In the present analysis, this approach is extended by dividing the tuning range into a set of subintervals $\C{\eta_k^c,\eta_{k+1}^c},\ k=1,\ldots,p-1$ and applying linearization (\ref{e8}) to each subinterval. The \emph{sampling range} $\C{\eta_1^c,\eta_p^c}$ is considered long enough to contain all the $\eta_i$ values attained during the array performance. In the first place, the characteristic of each $i-$th VCO is sampled under free-running conditions in circuit-level HB at a set of $p$ values $\eta_{1}^c<\cdots<\eta_{p}^c$ of the tuning voltage $\eta_i$, obtaining the set of steady state solutions $\LL{V_i^{o}(\eta_k^c),\omega(\eta_k^c),\eta_k^c}_{k=1}^p$. This calculation is made through finite differences in circuit-level HB using an AG, as explained in \cite{suarez:OA:noise,suarez:OA:general,suarez:libro2}. Then, the function $Y_i$ is considered as piecewise linear in the intervals $\C{\eta_k^c,\eta_{k+1}^c}$. For $\eta_i\in\C{\eta_k^c,\eta_{k+1}^c}$, the admittance function is approached by a first-order Taylor series about the nearest free-running solution in the sampling range $\Pa{V_i^{o}(\eta_k^c),\omega_i^o(\eta_k^c),\eta_k^c}$:
\
\begin{IEEEeqnarray}{lll} 
	Y_i\Pa{V_i,\omega_i,\eta_i}\simeq
	Y_{iV}^{k}\Delta V_i+Y_{i\omega}^{k}\Delta\omega_i+
	Y_{i\eta}^{k}(\eta_i-\eta_k^c),\nonumber\\
	\Delta V_i=V_i-V_i^{o}(\eta_k^c),\qquad\Delta\omega_i=\omega_i-\omega_i^{o}(\eta_k^c),\label{e2}\\
	I_{iG^{r,i}}(V_i,\omega_i,\eta_i)\simeq
	I_{iG^{r,i}}(V_i^o(\eta_k^c),\omega_i^o(\eta_k^c),\eta_k^c)\equiv I_{iG^{r,i}}^k,\nonumber\\
	Y_{ix}^k\equiv\D\frac{\Dp Y_i\Pa{V_i^{o}(\eta_k^c),\omega_i^{o}(\eta_k^c),\eta_k^c}}{\Dp x},\quad x\equiv V_i,\omega_i,\eta_i\nonumber
\end{IEEEeqnarray}

In the following, the new formulation (\ref{e2}) will be called piecewise (PW) SAF, to be distinguished from the previous non-PW SAF (\ref{e8}). 

\begin{figure}[!h] 
\centering
 \subfloat[]{\includegraphics[width=3in]{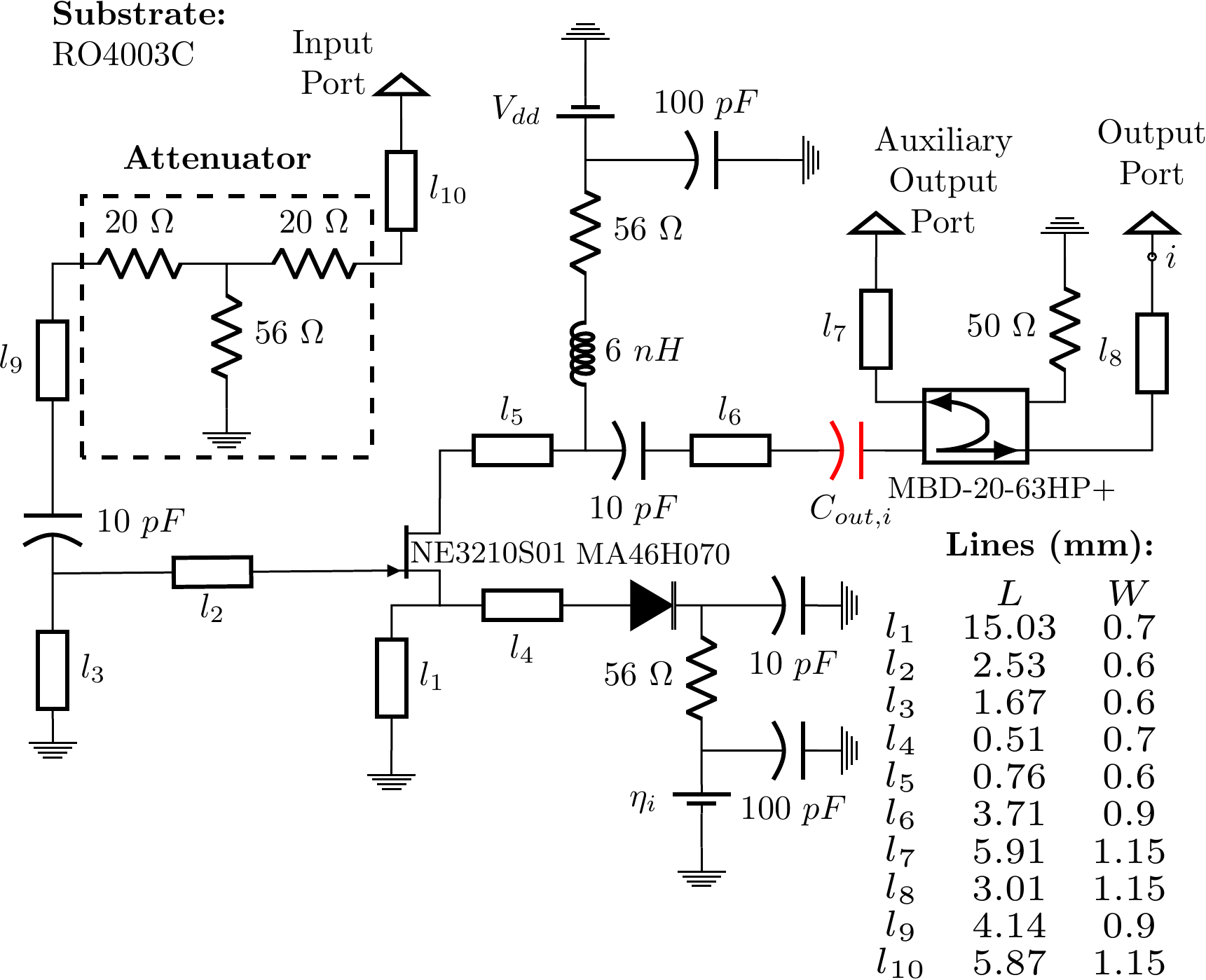}\label{f4a}}
\hfill\\
 \subfloat[]{\includegraphics[width=3in]{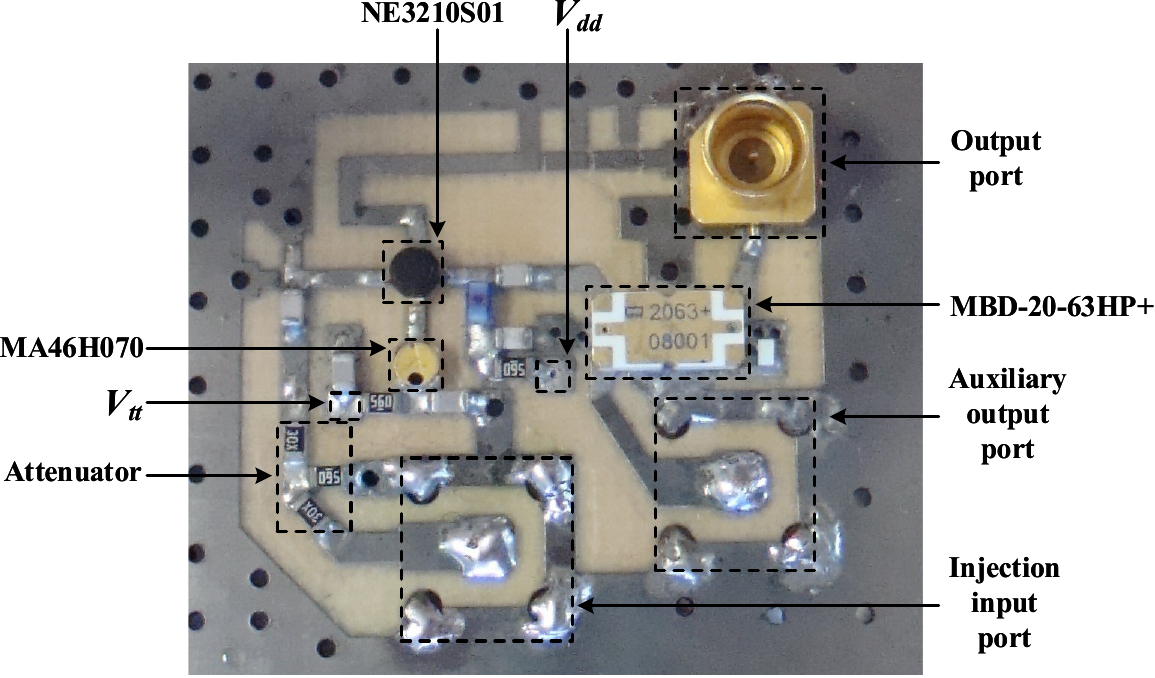}\label{f4b}}\\
\hfill\\
\caption{(a) Schematic of the individual $i-$th VCO of the array, where $V_{dd}$ and $\eta_i$ are the drain and tuning voltages, respectively. The variables $L$ and $W$ represent the length and the width of each line in $mm$. (b) Photograph of the individual VCO.}
\label{f4}
\end{figure}

The improvement in the accuracy achieved by the PW SAF has been illustrated by their application to the FET-based oscillator of Fig. \ref{f4}. This oscillator represents each individual $i-$th VCO constituting the array that will be analyzed in Section \ref{s4}. In Fig. \subref*{f4a}, the VCO schematic is shown, specifying all the necessary components to connect the oscillator to the network. This oscillator is designed using a field-effect transistor NE3210S01 on a RO4003C substrate. A varactor diode MA46H070 is used as a tuning element. The VCO has an input port that allows the injection of an external signal in the system. In order to illustrate the effect of parasitic components that may appear during the manufacturing process, a capacitor $C_{out,i}$ has been placed in the output path. The capacitance value $C_{out,i}$ can be different for each oscillator, modeling the possible differences between the oscillators resulting from this process.

\begin{figure}[!h] 
\fontsize{8}{10}\selectfont
\centering 
 \subfloat[]{\includegraphics[width=3in]{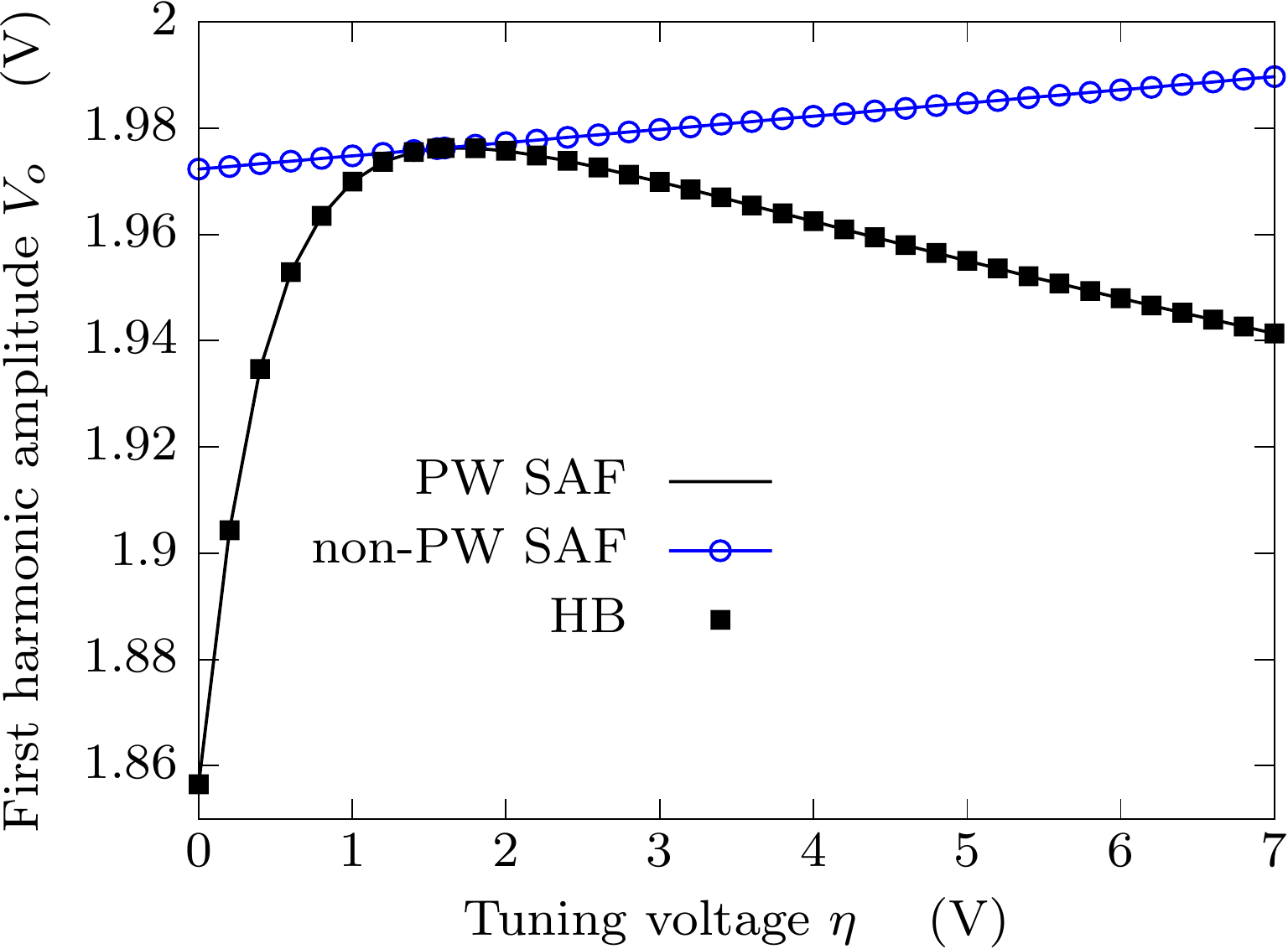}\label{f9a}}
\hfill\\
 \subfloat[]{\includegraphics[width=3in]{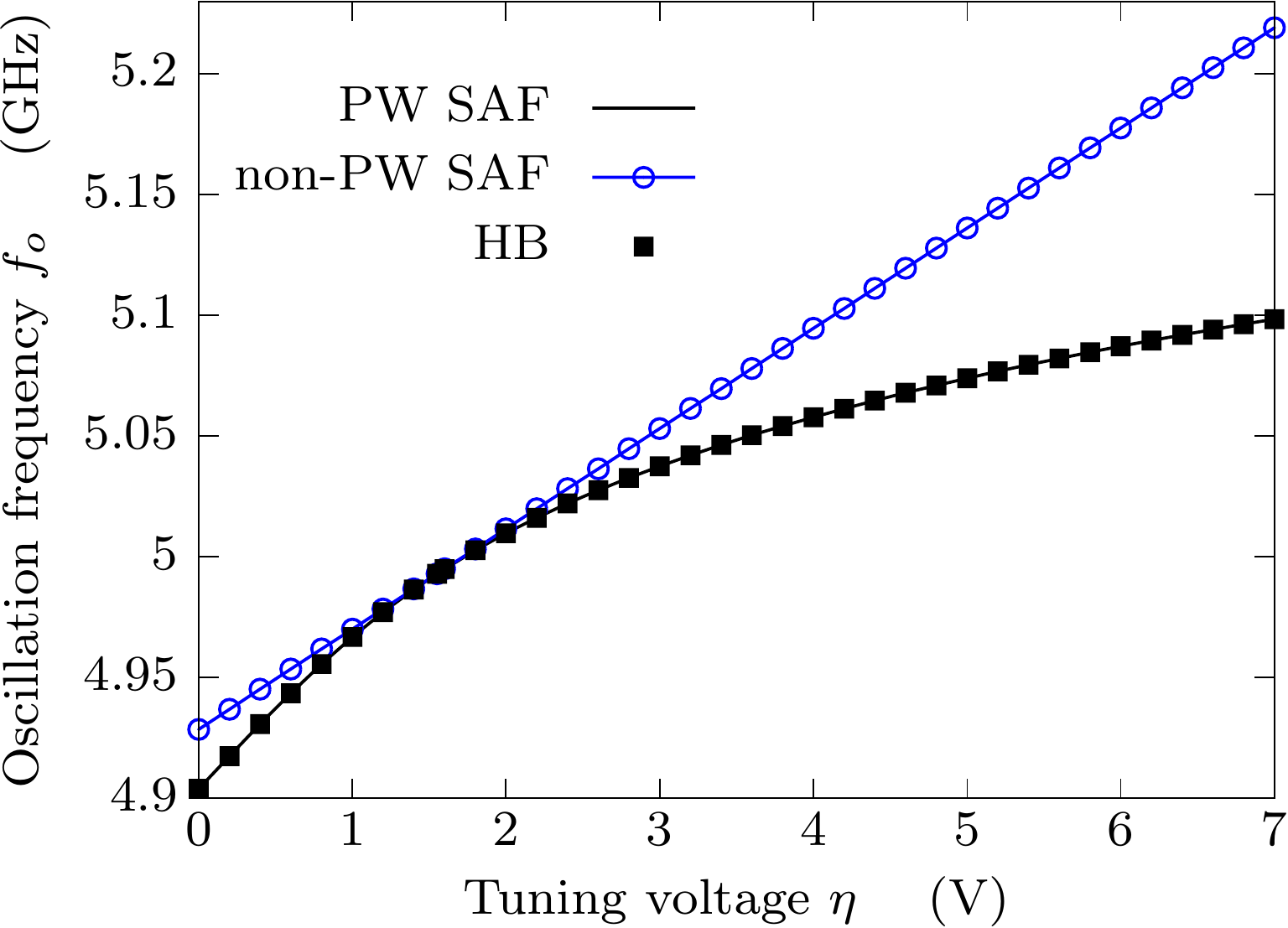}\label{f9b}}\\
\hfill\\
\caption{Evolution of the free-running steady state solution versus the tuning voltage $\eta$. Comparison of the results obtained with the non-PW SAF with the PW SAF. The PW SAF agrees with the circuit-level HB simulations at the points $\eta_i$ of the sampling interval. (a) First harmonic amplitude $V_o$. (b) Oscillation frequency $f_o$}
\label{f9}
\end{figure}

In Fig. \ref{f9}, the set of free-running steady-state solutions $\LL{V_i^o(\eta_i),\omega_i^o(\eta_i)}$ has been represented versus the tuning voltage $\eta_i$. These solutions have been obtained by solving the complex equation $Y_i(V_i,\omega_i,\eta_i)=0$ for each value of $\eta_i$ in the sampling range $\C{\eta_1^c,\eta_p^c}=\C{0,7}$ V. The results using formulations (\ref{e8}) and (\ref{e2}) have been compared and the solutions of the circuit-level HB simulations for each $\eta_k^c$ have been superimposed in both figures. As expected, the curves obtained through both formulations are tangent to each other at the tuning value $\eta^c=1{.}55\ V$, for which the non-PW SAF (\ref{e8}) has been derived. For $k=9$, at $\eta_9^c\simeq\eta^c$, equations (\ref{e8}) and (\ref{e2}) provide similar values for the functions $Y_i(V_i,\omega_i,\eta_i)$ and $I_{iG^{r,i}}(V_i,\omega_i,\eta_i)$, since $Y_{ix}^c\simeq Y_{ix}^9$ and $I_{iG^{r,i}}^c\simeq I_{iG^{r,i}}^9$. As a consequence, in a small interval of the tuning voltage about $\eta^c$, the current function $I_1^i$ given in (\ref{e3}) to model the VCO performance provides similar results when using both formulations. Fig. \ref{f9} shows that, as $\eta_i$ is shifted away from this value, the steady-state solutions $(V_i^o(\eta_i),\omega_i^o(\eta_i))$ distance from the values at $\eta^c$, making the derivatives $Y_{ix}^k$ diverge from $Y_{ix}$. Then, for these $\eta_i$ values, the PW SAF (\ref{e2}) provides a different and more accurate model than the non-PW SAF (\ref{e8}). Note that, in both figures, the PW SAF solutions are validated through comparison with circuit-level HB simulations at the points $\eta_k^c,\ k=1,\ldots,p$. In each interval $\C{\eta_k^c,\eta_{k+1}^c}$, the PW SAF assumes the function $I_1^i$ to be linear in the $(V_i,\omega_i,\eta_i,G^r,G^i)$ variables, and nonlinear in the phase variable $\phi_i$. Therefore, the samples $\eta_k^c,\ k=1,\ldots,p$ must be close enough each other for this approach to be valid.

\subsection{Formulation of the coupled-oscillator system} 
\begin{figure} [!h] 
    \centering
\includegraphics[width=3in]{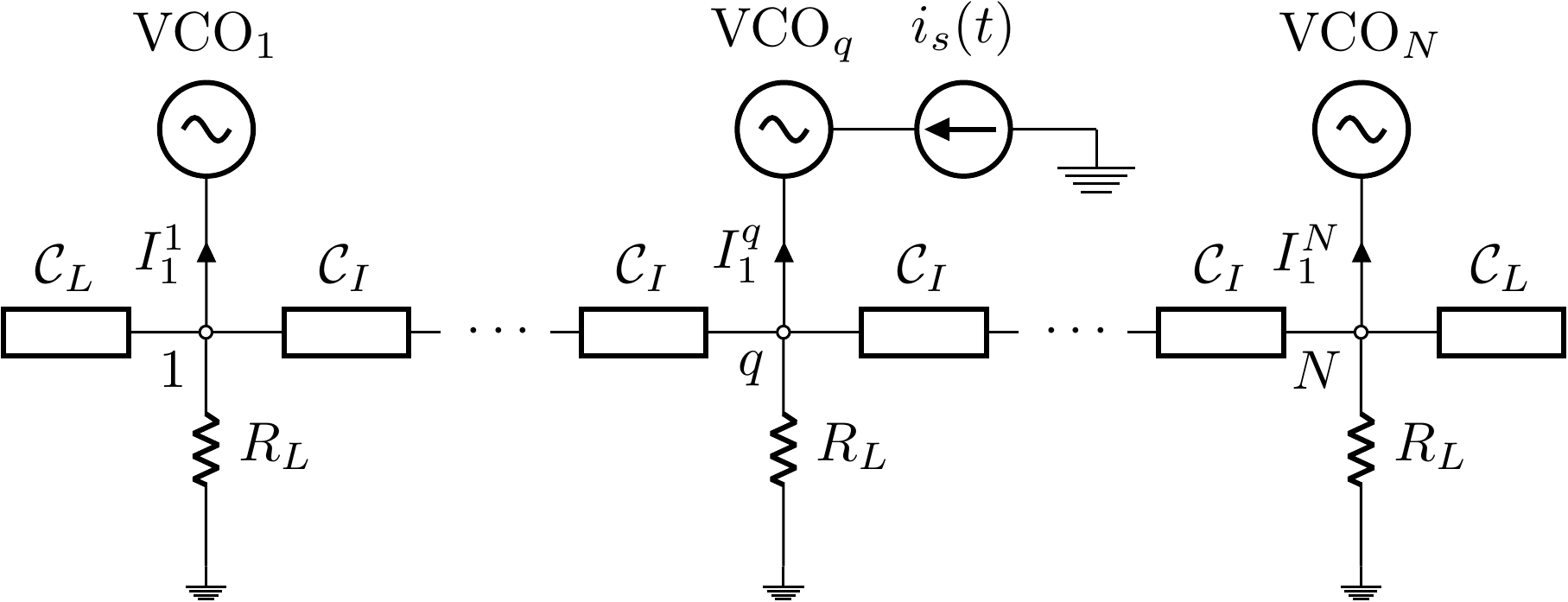}
 \caption{Schematic of the coupled oscillator array. The output port of each oscillator element is loaded with a resistor $R_L=50\ \Omega$ and coupled to the neighbor elements through a linear network.}
  \label{f1}
\end{figure}

The schematic of the coupled oscillator system is shown in Fig. \ref{f1}. The output port of each oscillator element is loaded with a resistor $R_L=50\ \Omega$ and coupled to the neighbor elements through a linear network $\mathcal{C}_I$, described by an $(2\times 2)$ admittance matrix \cite{Lynch:CO,Heath:beam}. In order to preserve the system symmetry, an additional monopole $\mathcal{C}_L$ providing an admittance reflection component $Y_s$ has been introduced to each edge oscillator. The coupled oscillator system is injected by an input generator, modeled by a current source $i_s(t)=I_s\cos(\omega_st+\theta_s)$ connected to the input port of the $q-$th oscillator of the array. Each $i-$th VCO is tuned by a varactor diode placed in the resonant network biased by a tuning voltage $\eta_i$, as shown in Fig. \ref{f4}. These elements are used in the array performance to control the phase shift between the VCOs \cite{suarez:OA:noise,suarez:OA:general,VCO_Array_7,VCO_Array_1}.

The formulation will be developed for the general case of an array with $N$ synchronized oscillators, fulfilling $\omega_i=\omega_s,\ i=1,\ldots,N$. By Kirchhoff's current law (KCL), the total current at the output node of each $i-$th oscillator must fulfill $i_{T,i}(t)=0,\ \forall t,i$. Translating the equation of each VCO to the frequency domain the following system is obtained \cite{suarez:OA:general}:
\
\begin{IEEEeqnarray}{lll}
	0=I_1^i(V_i,\phi_i,\omega_s,\eta_i,G^r,G^i)+\sum_{n=1}^NC_{i,k}(\omega_s)V_ke^{j\phi_k},\nonumber\\
	C_{i,k}(\omega)=\left\{\begin{array}{cl}
		Y_{nb},& |k-i|=1,\\
		Y_s,& k=i,\\
		0,& |k-i|>1
	\end{array}\right.  \label{e1}
\end{IEEEeqnarray}

where $i=1,\ldots N$. System (\ref{e1}) contains the set of  first harmonic KCL equations evaluated at the output node of each VCO, with $I_1^i$ and the summation term corresponding to the currents entering the VCO and the coupling network, respectively \cite{suarez:OA:noise,suarez:OA:general}. The PW SAF applied to the coupled system is derived as follows. Let the set of tuning voltages belong to the intervals $\eta_i\in[\eta^c_{k_i},\eta^c_{k_i+1}),\ i=1,\ldots,N$. Then, the admittance functions $Y_i$ are expressed using approach (\ref{e2}) and system (\ref{e1}) becomes:
\
\begin{IEEEeqnarray}{lll}
	0=\Pa{Y_{iV}^{k_i}\Delta V_i+Y_{i\omega}^{k_i}\Delta\omega+
	Y_{i\eta}^{k_i}(\eta_i-\eta^c_{k_i})}V_i+\nonumber\\
	+\sum_{n=1}^NC_{i,n}(\omega_s)
	V_ke^{j(\phi_k-\phi_i)}+\label{e4}\\
	+\delta_i^qI_s\Pa{I_{iG^r}^{k_i}\cos(\theta_s-\phi_i)+
	I_{iG^i}^{k_i}\sin(\theta_s-\phi_i)}\nonumber
\end{IEEEeqnarray}

where $i=1,\ldots,N$ and the Kronecker delta $\delta_i^q$ determines that the injection current source is connected to the $q-$th VCO of the array. To calculate the derivatives $\LL{Y_{ix}^{k_i},I_{iG^{r,i}}^{k_i}}$ for each pair $\Pa{i,k_i}$, first, the values $V_{i}^o(\eta^c_{{k_i}})$ and $\omega_{i}^o(\eta^c_{{k_i}})$ corresponding to the free-running $i-$th VCO are obtained through circuit level HB simulations as described in Section \ref{s3}. Then, the derivatives are calculated using the definitions given in (\ref{e2}).  This procedure is repeated for $i=1,\ldots,N$ and $k_i=1,\ldots,p$ in order to prepare system (\ref{e4}) to cover all the possible scenarios of the array performance. 

\subsection{System resolution} \label{s2}
System (\ref{e4}) is composed by $N$ complex equations with $3N+1$ unknowns $\LL{V_i,\phi_i,\eta_i,\omega_s},\ i=1,\ldots,N$. The vector of coefficients $\V{k}=\C{k_1,\ldots,k_N}^t$ indexing the set of derivatives $\LL{Y_{ix}^{k_i},I_{iG^{r,i}}^{k_i}}_{i=1}^N$ is determined by the sampling range subintervals containing the tuning voltage values $\eta_i\in[\eta^c_{{k_i}},\eta^c_{k_i+1})$. To balance this system, $N+1$ additional real equations are required, which are chosen taking into account the array performance. For beam steering applications, there must be a constant phase shift $\Delta\varphi$ between the oscillator elements. To assure convergence to this constant phase-shift solution, we impose the condition $\phi_{i+1}-\phi_i=\Delta\varphi$, with $i=1,\ldots,N-1$. It is easy to see that system (\ref{e4}) is invariant versus a global phase shift $\phi_1+\alpha,\ldots,\phi_N+\alpha,\theta_s+\alpha$. Thus, for its practical resolution, the $q-$th VCO phase value $\phi_q$ can be arbitrarily set to zero. To complete the set of $N+1$ additional equations, the $q-$th VCO tuning voltage $\eta_q$ can be fixed to a specific value, which is chosen to maximize the available tuning range in the array. Taking this into account, system (\ref{e4}) can be rewritten in compact form as:
\
\begin{IEEEeqnarray}{lll}
	\V{H}\Pa{\V{V},\V{\eta},\omega_s,\Delta\varphi}=\V{0},\label{e5}\\
	\V{V}=\C{V_1,\ldots,V_N}^t,\quad
	\V{\eta}=\C{\eta_1,\ldots,\eta_{q-1},\eta_{q+1},\ldots,\eta_N}^t\nonumber
\end{IEEEeqnarray}

where the $N$ components of the complex vector function $\V{H}$ are given by the right-hand parts of the equations in (\ref{e4}). When the constant phase shift $\Delta\varphi$ is fixed to a given value, system (\ref{e5}) gets determined and the set of unknowns $\LL{\V{V},\V{\eta},\omega_s}$ can be solved with any optimization technique. At each step of the optimization procedure, the vector of coefficients $\V{k}$ must be actualized according to the subintervals containing  the set of tuning values $\eta_i\in[\eta^c_{{k_i}},\eta^c_{k_i+1}),\ i=1,\ldots,N$. 

{ The presented formulation provides a general description of the coupled oscillator system. Indeed, in the general case of $N$ oscillators, for a given constant phase shift value $\Delta\varphi$, system (\ref{e5}) must be fulfilled for the set of unknowns $\LL{\V{V},\V{\eta},\omega_s}$, where $\V{\eta}=\C{\eta_1,\ldots,\eta_{q-1},\eta_{q+1},\ldots,\eta_N}^t$ contains the tuning voltages of all VCOs except $\eta_q$, corresponding to the VCO with phase $\phi_q=0$. In the particular case of $N=3$ oscillators, the set of tuning voltages to be solved gets reduced to $\eta_1$ and $\eta_3$, corresponding to the outermost VCOs. It must be noted that, as demonstrated in \cite{york:quiasioptical}, for $N>3$, the inner tuning voltages of the VCOs $i=2,\ldots,N-1$ remain nearly invariant in the constant phase-shift interval.} Note that the ability of the injected VCO to lock on the injection signal depends on the values of the array derivatives and the coupling network, which must provide a real and consistent solution for the set of unknowns.

\subsection{Stability analysis} \label{s1}
The stability of the synchronized solutions can be determined by applying a perturbation analysis to the PW SAF (\ref{e4}). In general, the stability of a given synchronized solution $\LL{V_i,\phi_i,\eta_i,\omega_s},\ i=1,\ldots,N$ is analyzed by introducing time-varying perturbation components in the amplitude and phase variables \cite{suarez:OA:noise,suarez:OA:general}:
\
\begin{equation}
	V_i\mapsto V_i+\Delta v_i(t),\ 
	\phi_i\mapsto \phi_i+\Delta\phi_i(t),\quad i=1,\ldots,N
\end{equation}

Then, translating the frequency-domain system (\ref{e4}) to the envelope domain as described in \cite{kurokawa:free} and canceling the steady state terms, the following linear time-invariant (LTI) system is obtained:
\
\begin{IEEEeqnarray}{lll}
	0=Y_{iV}^{k_i}\Delta v_i+Y_{i\omega}^{k_i}
	\Pa{\Delta\dot{\phi}_i-j\frac{\Delta\dot{v}_i}{V_i}}+\label{e6}\\
	+\frac{jV_k}{V_i}\sum_{n=1}^NC_{i,n}(\omega_s)
	e^{j(\phi_k-\phi_i)}
	\Pa{\Delta\phi_k-\Delta\phi_i-j\frac{\Delta v_k}{V_k}}+\nonumber\\
	+\delta_i^qI_s\Pa{I_{iG^r}^{k_i}\sin(\theta_s-\phi_i)-
	I_{iG^i}^{k_i}\cos(\theta_s-\phi_i)}\Delta\phi_i\nonumber
\end{IEEEeqnarray}

with $i=1,\ldots,N$ and $\eta_i\in[\eta^c_{k_i},\eta^c_{k_i+1})$, and where high order terms in the perturbation components have been neglected. Since, as explained in Section \ref{s2}, the synchronized solution $\LL{V_i,\phi_i,\eta_i,\omega_s},\ i=1,\ldots,N$ and the vector $\V{k}$ of derivative coefficients are determined by the constant phase shift value $\Delta\varphi$, system (\ref{e6}) can be rewritten in compact form as:
\
\begin{equation}
\begin{aligned}
	&\Delta\dot{\V{X}}=A\Pa{\Delta\varphi}\Delta\V{X}
\end{aligned}
\end{equation}

with $\Delta\V{X}=\C{\Delta v_1,\ldots\Delta v_N,\Delta\phi_1,\ldots,\Delta\phi_N}^t$. The stability analysis is carried out in the following way. In the first place, the synchronized solution corresponding to a given constant phase shift $\Delta\varphi$ is obtained by solving system (\ref{e5}). The matrix $A\Pa{\Delta\varphi}$ is constructed from this solution. Then, the set of eigenvalues $\LL{\lambda_1,\ldots,\lambda_{2N}}$ of matrix $A\Pa{\Delta\varphi}$, which are called the \emph{system poles}, are calculated. The \emph{stable range} is the range of $\Delta\varphi$ values for which $\max\LL{\operatorname{Re}(\lambda_i)}_{i=1}^{2N}<0$.

\section{Application to the steady-state analysis of the coupled-oscillator system}
\subsection{Study of the effect of asymmetries in an array of Van der Pol-type oscillators}
\begin{figure} [!h] 
    \centering
  \subfloat[]{\includegraphics[width=3in]{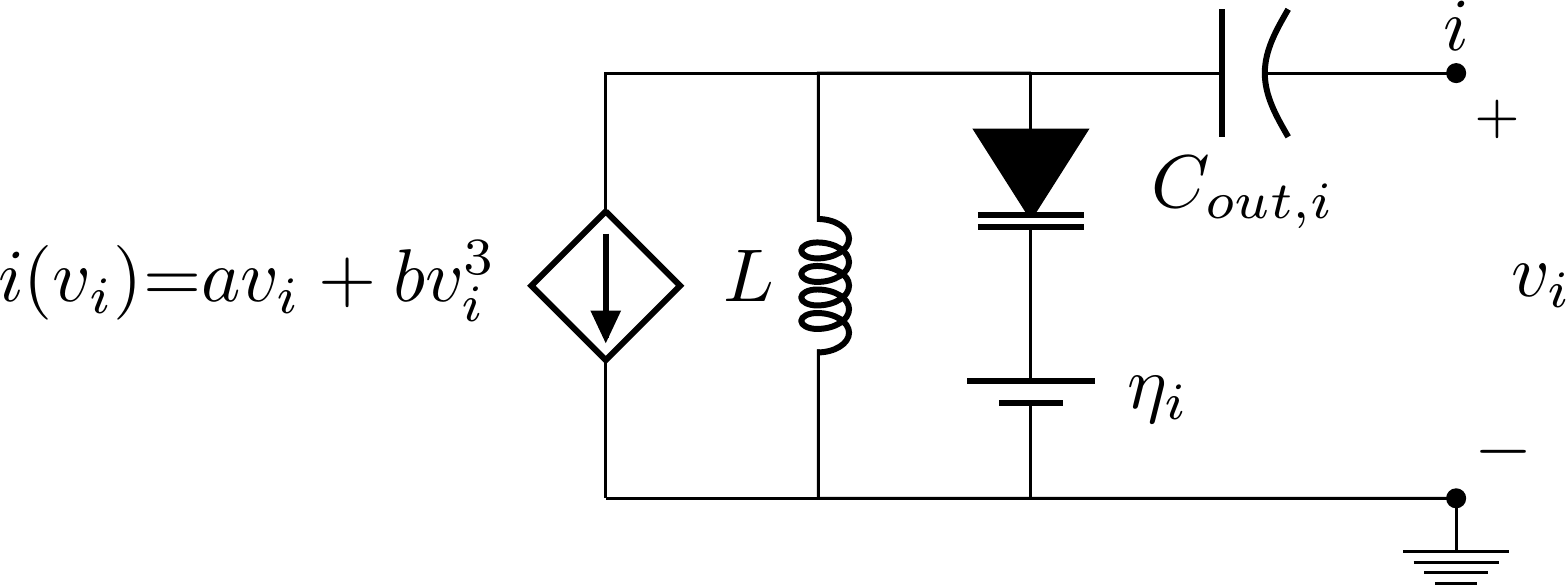}\label{f2a}}
\hfill\\
  \subfloat[]{\includegraphics[width=1.8in]{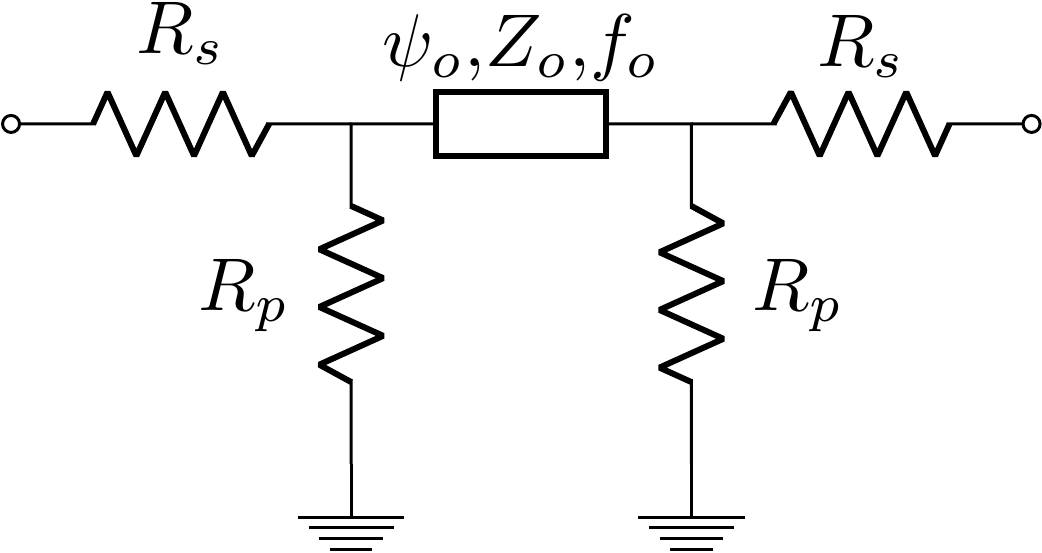}\label{f2b}}
    \hfill
  \subfloat[]{\includegraphics[width=1.3in]{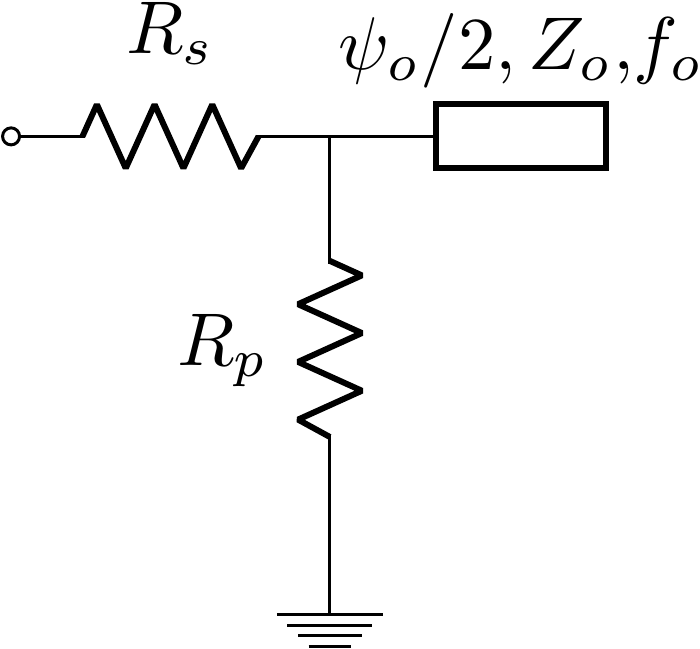}\label{f2c}}
  \caption{Components of the array of Van der Pol-type oscillators. (a) Schematic of each VCO, with $a=-0{.}023\ A/V$, $b=0{.}01\ A/V^3$ and $L=1{.}53\ nH$. The varicap is constituted by a p-n diode with $C_{jo}=0{.}72\ pF$ and $M={0.}5$. The capacitor $C_{out,i}$ models the parasitic components that may appear during the manufacturing process, producing differences between the oscillators in the array.  (b) Coupling network $\mathcal{C}_I$, containing an ideal transmission line with $\psi_o=360\ deg$, $Z_o=50\ \Omega$, $f_o=5{.}2\ GHz$, and the resistances $R_s=1250\ \Omega$, $R_p=300\ \Omega$ (c) Load $\mathcal{C}_L$ connected to each of the outermost oscillators preserving the system symmetry.}
  \label{f2} 
\end{figure}

As stated in the Introduction, the simulation of the coupled-oscillator system in circuit-level HB may present convergence issues which cannot be overcome even with the aid of auxiliary generators (AG) \cite{suarez:OA:noise,suarez:OA:general,suarez:libro2}. Then, in order to compare the presented technique with circuit-level HB simulations, it has been applied to a free-running array of $N=3$ Van der Pol-type oscillators, in the absence of injection source $(I_s=0)$. The simplicity of the involved devices makes this system suitable for this kind of simulation. The schematic of each $i-$th oscillator is shown in Fig. \subref*{f2a}. The oscillation is due to the performance of a nonlinear current source $i(v_i)$, acting as active device, in conjunction with a parallel $RLC$ resonator. The resonator capacitance is provided by a varactor diode, biased by a dc voltage source $\eta_i$ acting as a tuning voltage, whereas the resistance is given by the output load $R_L$ shown in Fig. \ref{f1}. 

Each $i-$th oscillator is connected to the array at the output node $i$. A capacitor $C_{out,i}$ has been introduced before the output port to model small differences between the oscillators in the array. The schematics of the linear networks $\mathcal{C}_I$ together with the symmetry components $\mathcal{C}_L$ are shown in Figs. \subref*{f2b} and \subref*{f2c}, respectively.  

\begin{figure}[!h] 
\fontsize{8}{10}\selectfont
\centering 
 \subfloat[]{\includegraphics[width=3in]{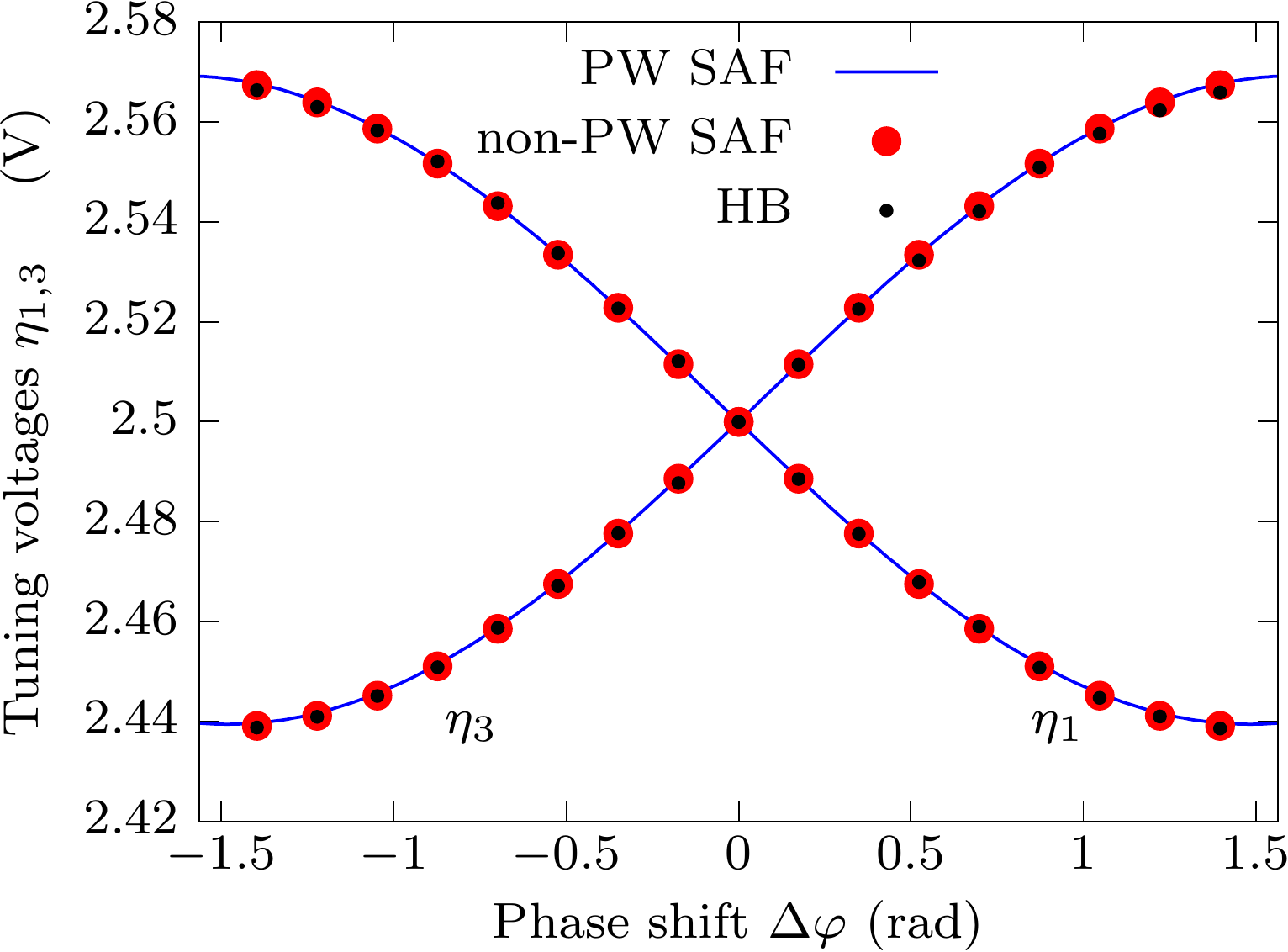}\label{f3a}}
\hfill\\
 \subfloat[]{\includegraphics[width=3in]{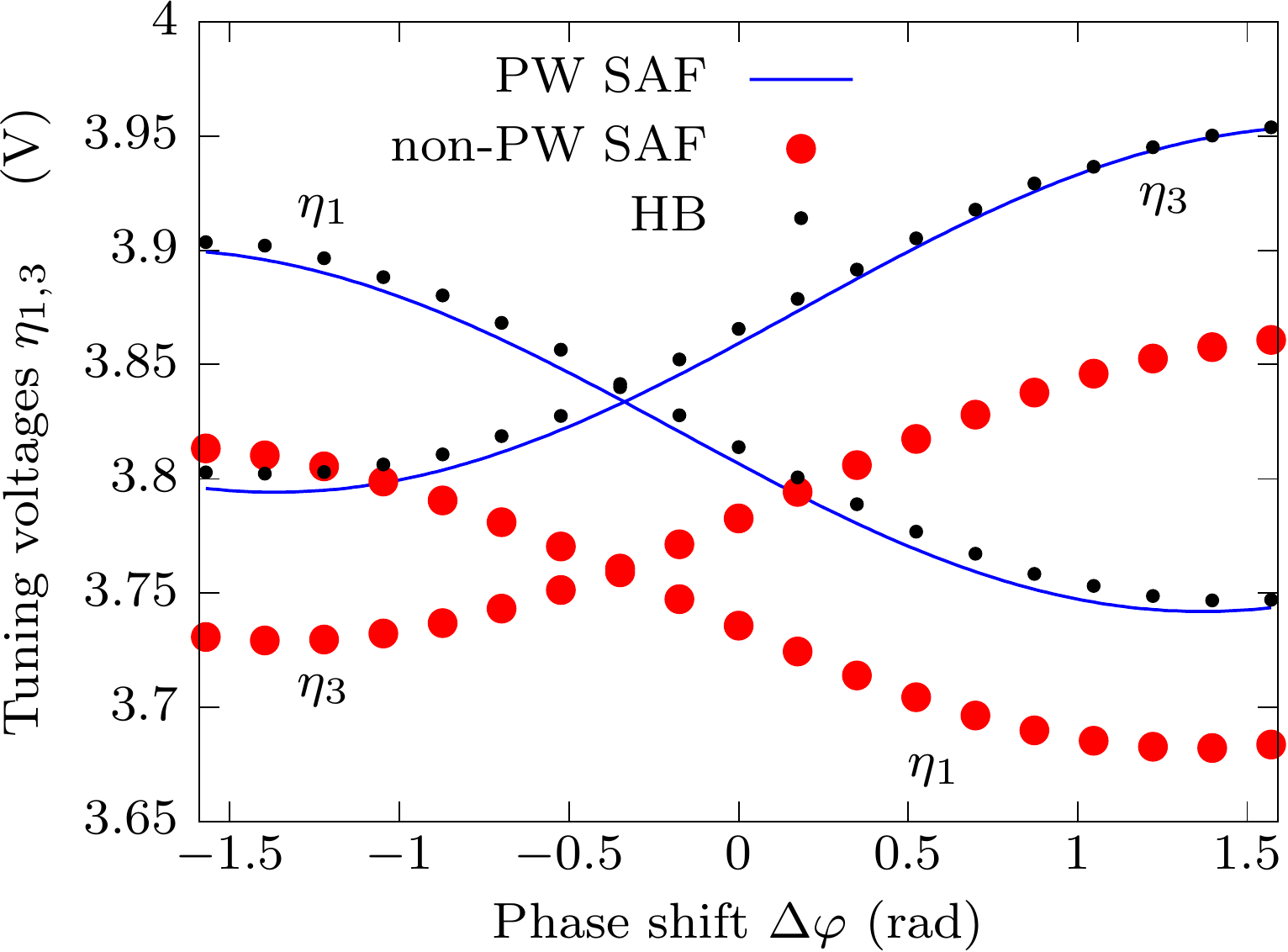}\label{f3b}}\\
\hfill\\
\caption{Array of $N=3$ Van der Pol-type oscillators. Simulation of the evolution of $\eta_{1,3}$ along the stable range of constant phase-shift solutions. (a) Identical oscillators. The tuning voltages vary symmetrically and the PW SAF with $p=33$ points, the non-PW SAF and the circuit-level HB techniques provide similar results. (b) Non-identical oscillators. The symmetry of the previous case is broken, and the non-PW SAF technique presents discrepancies with HB and PW SAF techniques.}
\label{f3}
\end{figure}

Fig. \ref{f3} shows the evolution of the outermost tuning voltages $\eta_{1,3}$ to obtain a constant phase-shift solution as $\Delta\varphi$ varies along the stable range. To balance system (\ref{e4}), the tuning voltage of the central oscillator has been fixed to the value $\eta_2=2{.}5\ V$, corresponding to the free-running frequency value $f_{o2}=5{.}2\ GHz$. In the PW SAF technique, a set of $p=33$ equispaced  samples in the sampling range $\C{2{.}4,4}\ V$ has been considered. Note that, since, in this case, there is not injection signal, the frequency unknown $\omega_s$ in (\ref{e5}) refers to the mutual synchronization frequency of the system.

In the first place, the case of three identical oscillators has been analyzed, removing the capacitors $C_{out,i}$. The poles $\LL{\lambda_i}_{i=1}^{6}$ have been calculated for each synchronized solution, indicating that the constant phase shift solutions are stable in the range $\Delta\varphi\in\Pa{-\pi/2,\pi/2}$. In Fig. \subref*{f3a} the simulation results provided by the PW SAF, the non-PW SAF and the circuit-level HB techniques are nearly identical for this case. As can be seen in this figure, the variation of $\eta_{1,3}$ is symmetric with respect to the in-phase solution $\Delta\varphi=0$. Due to the array symmetry, in order to achieve a particular phase shift value $\Delta\varphi$, the tuning voltages of the outermost oscillator elements must be shifted the same amount $\Delta\eta_1=-\Delta\eta_3=\Delta\eta(\Delta\varphi)$ in opposite directions from the central value $\eta_{1,3}=2{.}5\ V$, fulfilling $\Delta\eta(-\Delta\varphi)=-\Delta\eta(\Delta\varphi)$.

This symmetry is broken in the case of Fig. \subref*{f3b}, where the capacitors $C_{out,1}=10\ pF$ and $C_{out,3}=9{.}65\ pF$ have been introduced in the outermost oscillators of the array. In addition, the introduction of both capacitors shifts the range of variation of $\eta_{1,3}$ with respect to the symmetric case. As a consequence, the results of the non-PW SAF present discrepancies with the circuit-level HB simulation, since the former technique linearizes the VCO characteristic at the central VCO tuning value $\eta^c=2{.}5\ V$, far from the range of variation of $\eta_{1,3}$. In contrast, when using the PW SAF this range is more accurately predicted, since it is included in the considered sampling range. The purpose of this example is to show that, while the non-PW SAF lacks accuracy as the tuning voltage values $\eta_i$ are shifted away from the linearization value $\eta^c$, the PW SAF provides accurate solutions as long as the values $\eta_i$ lie inside the sampling range. In this particular case, the inaccuracies of the non-PW SAF appearing after the introduction of capacitors $C_{out,1,3}$ could be corrected by moving the linearization value $\eta^c$ of the outermost VCOs to the value $\eta^c\simeq 3{.}8$ V, lying in the middle of the  tuning range (see Fig. \subref*{f3b}). This correction becomes less effective as the tuning range gets higher, as it is the case of the coupled oscillator system analyzed in the next Section.

\subsection{Analysis of an array of FET-based oscillators} \label{s4}
The piecewise SAF has been applied to an array of $N=3$ FET-based oscillators, whose schematic is provided in  Fig. \ref{f4}. The varactor diode MA46H070 is used as a tuning element for the control of the phase shift. An auxiliary output port has been implemented through an integrated coupler to take a sample of the output voltage with the aim of measuring the phase shifts between the oscillators. The capacitor $C_{out,i}$ modeling small differences between the oscillators due to the manufacturing process has been introduced in VCOs $1,3$, with $C_{out,1}=3\ pF$ and $C_{out,3}=4\ pF$. The oscillators are coupled following the schematic of Fig. \ref{f1} with the networks $\mathcal{C}_I$ and $\mathcal{C}_L$ represented in Figs. \subref*{f2b} and \subref*{f2c}, using resistively loaded microstrip lines of length and width $(W,L)$ and $(W,L/2)$, respectively, with $W=1{.}85\ mm$ and $L=27{.}5\ mm$, and the resistors $R_s=120\ \Omega$, $R_p=220\ \Omega$.

For a better understanding of the system and the measuring procedure, in Fig. \ref{f5} the schematic of the measurement setup has been shown. The oscillator array has been separated into two boards: one with the three VCOs and other with the coupled network. A third board has been designed in order to develop the phase measurement. This circuit, made with two $360\ deg$ phase detector cells \cite{arana:measurement}, takes a sample of the outputs of the VCOs through their \emph{auxiliary output ports} through an integrated coupler to determine the phase shift between them. All the injection ports are loaded with $R_{inj}=50\ \Omega$ when the system works in free-running regime. In the injected regime this resistor is replaced by a signal generator in VCO $2$.  The resistors $R_{L}=50\ \Omega$ in VCOs $1,3$ and the spectrum analyzer input impedance in VCO $2$ take the place corresponding to the antenna array. To control the system, a Matlab script that communicates with the hardware has been developed. This script establishes the control voltages of the array and determines the phase shift between the output voltages of the VCOs from the signals provided by the phase measurement system. The synchronized state of the oscillators can be checked through the spectrum analyzer.

\begin{figure}[!h] 
\centering
\includegraphics[width=3in]{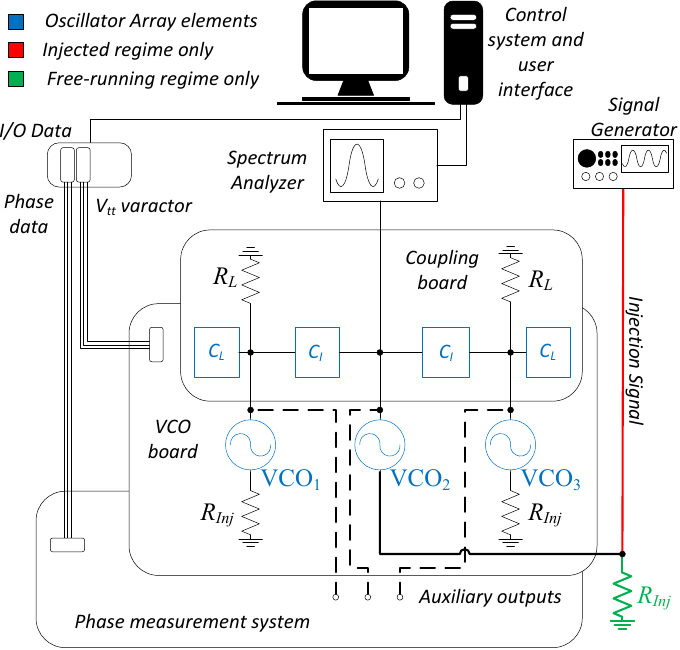}\label{f5}\\
\hfill\\
\caption{Measurement setup}
\label{f5}
\end{figure}

The PW SAF simulation technique has been tested by analyzing the array performance both in free-running and injected regimes. For this purpose, the set of derivatives $\LL{Y_{ix}^{k_i},I_{iG^{r,i}}^{k_i}}_{i=1}^N$ has been calculated from each free-running VCO characteristic with $p=27$ points in the sampling range $[0,5]\ V$, using circuit-level HB simulations in commercial software. To balance system (\ref{e4}) the tuning voltage of the central VCO has been fixed to $\eta_2=1{.}55\ V$. In this array, although the circuit-level HB technique with the aid of an auxiliary generator (AG) is able to simulate the free-running VCO steady state, it is unable to converge when applied to the whole array. For this reason, in the following analyses the simulation results of the PW SAF technique have been validated by comparison with measurements.

{ In the following, the main difficulty for the harmonic balance (HB) technique to simulate the system of coupled oscillators is explained. The autonomous oscillating solution of a VCO always coexists with a non-oscillating one. In the autonomous oscillating solution of a system of coupled oscillators, the oscillating regime of each one is excited. Then, the HB optimization technique must be able to reach the oscillating solution of each VCO in the coupled system. Some commercial HB simulators implement a single OscPort for the oscillator simulation. { The Keysight Advanced Design System (ADS) software uses OscPort components to simulate the oscillator steady state solution in HB. As explained in \cite{keysight:oscport}, the OscPort component is inserted either in the feedback loop of the oscillator, or between the parts of the circuits that have negative resistance and the resonator. Then, in the HB simulation, this component aids the simulator to calculate the amplitude and frequency of the steady state oscillation.}
 Although this is a useful technique for the single VCO simulation, it is in general unable to excite the oscillating solution of all the VCOs in the system. To circumvent this limitation, in \cite{collado:coupled:HB}, the use of $N$ auxiliary generators (AG), each one connected to a VCO in the array, has been proposed to prevent the HB optimization procedure from converging to the non-oscillating solution. The amplitude, phase and frequency of each AG become part of the system unknowns. Each AG contains an ideal bandpass filter that blocks all the harmonic components except from the first one. In this simulation, the HB system must be solved in combination with $N$ additional constraints. Each constraint is given by the complex equation $I_{AG}^k=0,\ k=1,\ldots,N$, where $I_{AG}^k$ is the first harmonic component of the steady current entering the AG connected to the $k-$th VCO. As the VCOs topologies become complex, the system asymmetries, produced by the different steady state solutions of each VCO, make it very difficult to solve the global system of equations (including the constraints), and this technique becomes inaccurate. This is the case of the present system, where the asymmetries arising as the constant phase shift between oscillators is swept make the HB optimizer unable to find the oscillating solution in the space of variables. In the array of $N=3$ oscillators considered in this paper, neither the use of AGs nor the OscPort technique have been able to converge to the system oscillating solution. When using the AG technique, three auxiliary generators have been used, each one connected to a VCO in the array, without success.}

\subsubsection{Analysis of the free-running array}
\begin{figure}[!h] 
\fontsize{8}{10}\selectfont
\centering 
 \subfloat[]{\includegraphics[width=3in]{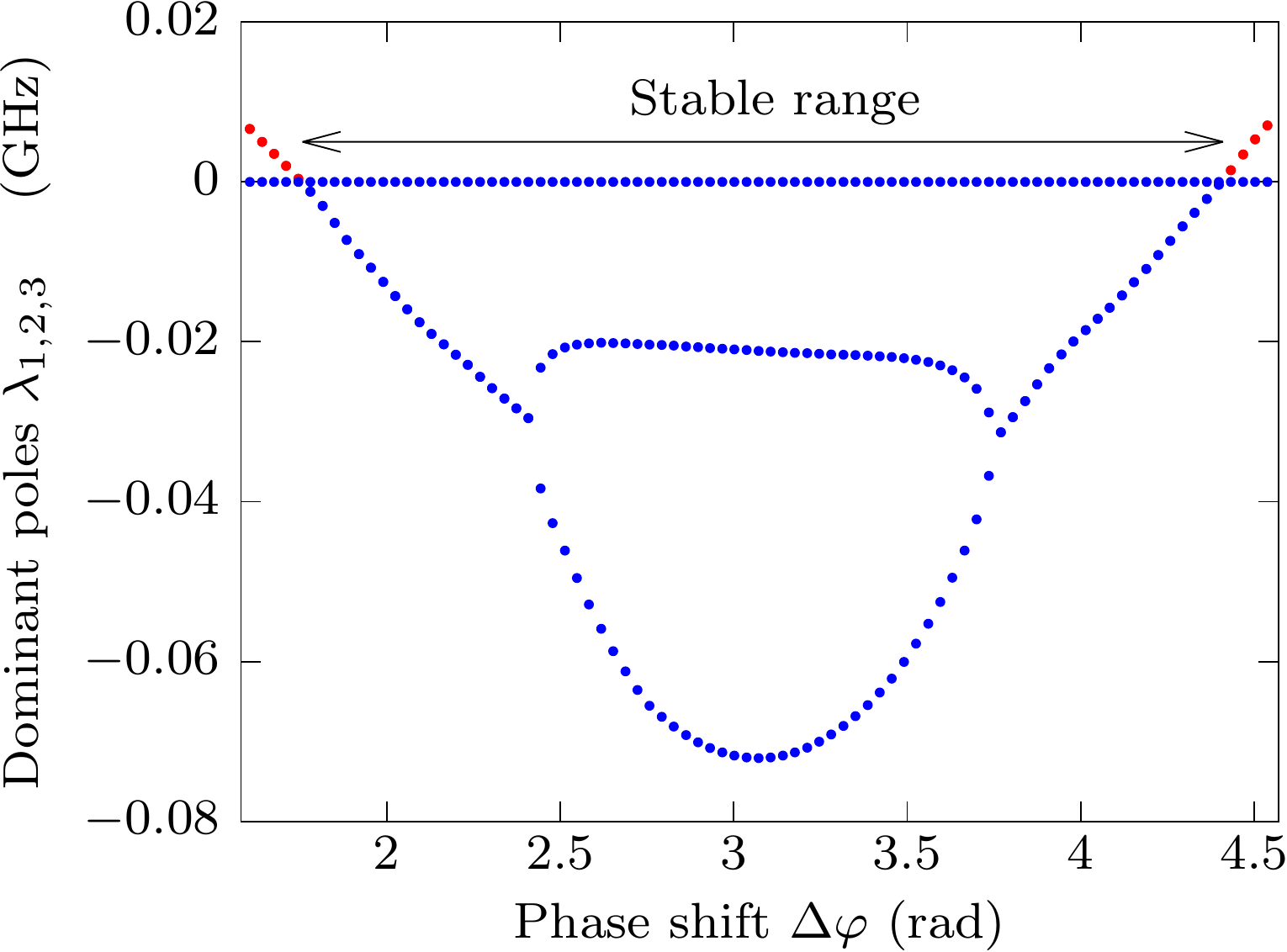}\label{f6a}}
\hfill\\
 \subfloat[]{\includegraphics[width=3in]{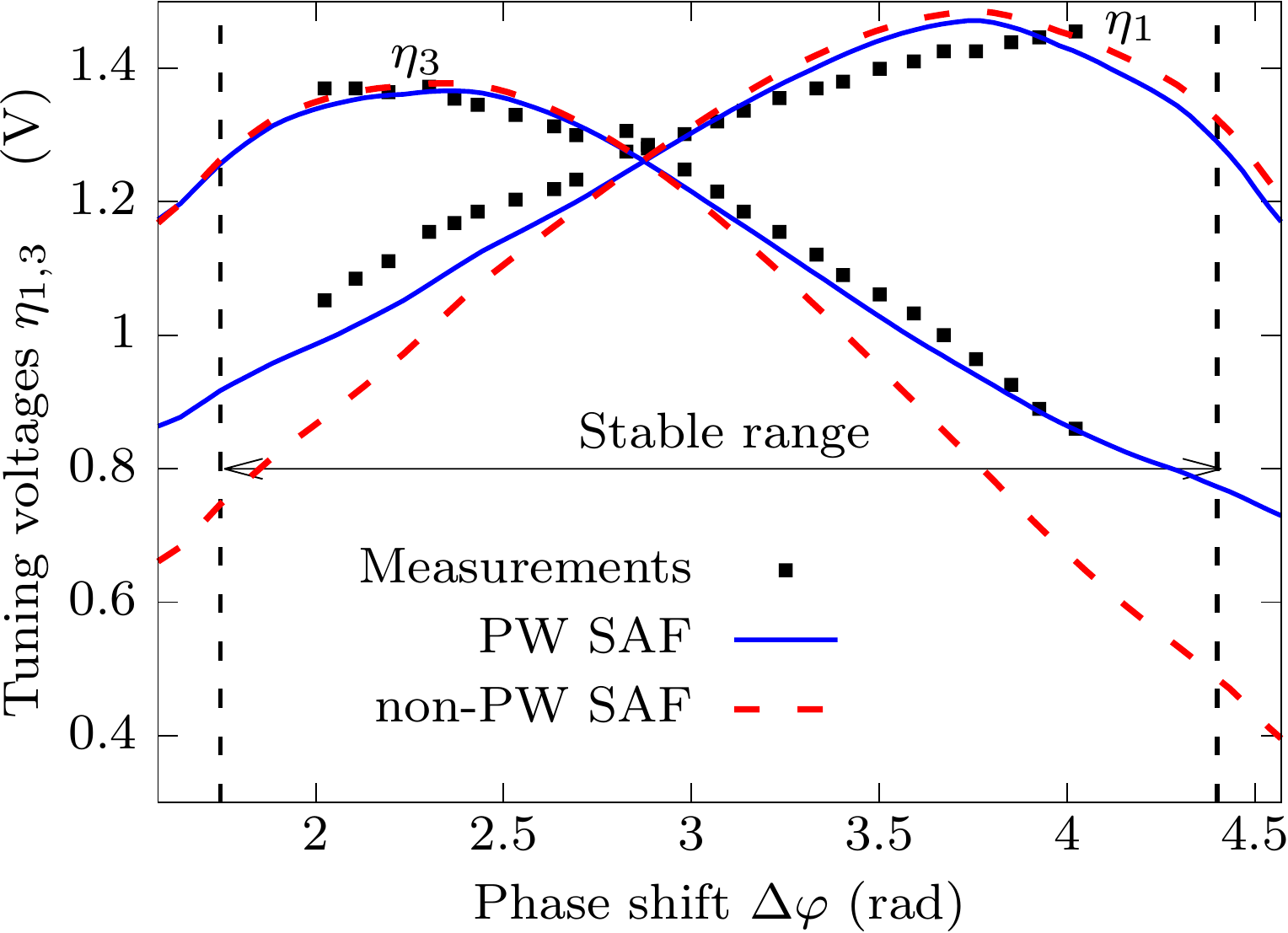}\label{f6b}}\\
\hfill\\
\caption{Free-running array. (a) Evolution of the dominant poles $\lambda_{1,2,3}$ with the constant phase shift $\Delta\varphi$, determining the stable range. (b) Variation of the tuning voltages $\eta_{1,3}$ versus $\Delta\varphi$.}
\label{f6}
\end{figure}

In this case, the technique has been tested by predicting the synchronized behavior of the free-running array, with $I_s=0$. The set of unknowns $\LL{\V{V},\V{\eta},\omega_s}$ has been calculated from system (\ref{e5}) as the constant phase shift $\Delta\varphi$ is swept in the range $[0,2\pi]$. { Systems of coupled oscillators can present several steady state solutions coexisting with the constant phase shift one. As it was shown in \cite{suarez:OA:noise}, the SAF can track the path of constant phase shift solutions with low computational cost. The reason for this is that, under weak coupling conditions, in the constant phase shift regime, the currents introduced by the coupling network give rise to a relatively small variation of the original free-running regime of the oscillator elements \cite{Lynch:CO,Heath:beam}. Then, since the SAF departs from the free-running regime of each oscillator, it converges easily to the constant phase shift solutions. In order to ensure that a solution provided by the SAF is physically observable, a stability analysis must be carried out.} For each $\Delta\varphi$ value, the stability of the synchronized solution has been analyzed following the perturbation analysis described in Section \ref{s1}. The resulting dominant poles $\lambda_{1,2,3}$ have been represented in Fig. \subref*{f6a}. In this case, the length of the transmission line of the coupling network produces a stable region centered at $\Delta\varphi=\pi\ rad$. 

In Fig \subref*{f6b}, the variation of the outermost VCOs tuning voltages $\eta_{1,3}$ in the stable region has been shown. Note that this variation is asymmetrical due to the slightly different admittance derivatives of each oscillator, produced by the capacitors $C_{out,i}$. In this analysis, the mutual synchronization frequency keeps nearly constant with slight variations at $f_s\simeq 5\ GHz$. In the same figure, the results provided by the non-PW SAF have been superimposed. As can be seen, the PW SAF provides a better agreement with measurements than the non-PW SAF. As expected, this occurs specially in the regions where the tuning voltage values are farther from the point $\eta=1{.}55\ V$ about which the first order Taylor series has been calculated for the non-PW SAF.

\begin{figure} [!h] 
\fontsize{8}{10}\selectfont
    \centering
\includegraphics[width=3in]{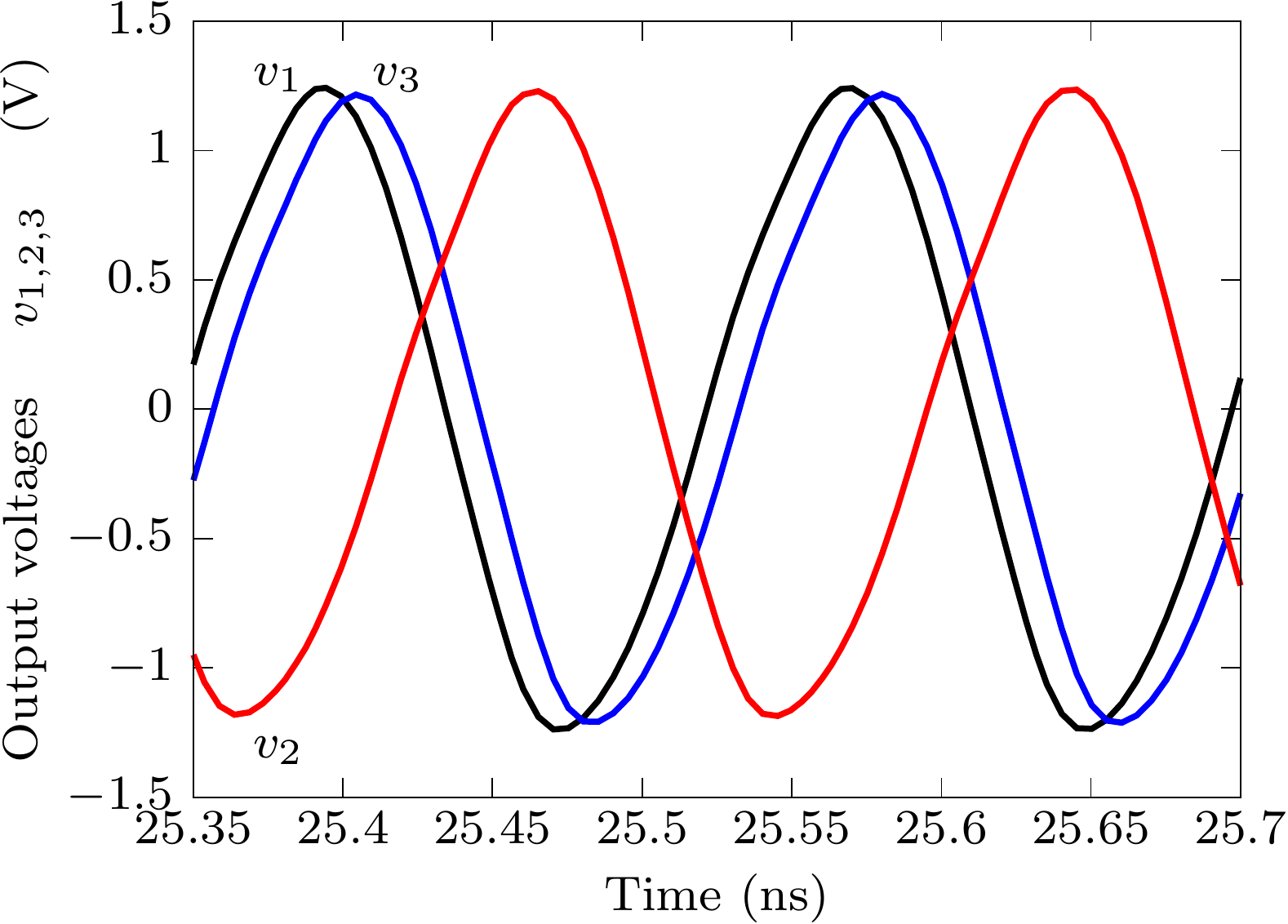}
 \caption{{ Free-running array. Transient simulation of the output voltage waveforms. The tuning voltages $\eta_{1,2,3}$ have been set to the values provided by the SAF for $\Delta\varphi=\pi$ rad.}} \label{f10}
\end{figure}

Once the SAF has provided the tuning voltages required to obtain a certain phase shift $\Delta\varphi$ value in the free-running array, the transient simulation { using Keysight Advanced Design System (ADS) simulator} has been used to observe the output voltage waveforms corresponding to these tuning values. Fig. \ref{f10} shows the circuit-level transient simulation of the output voltage waveforms. In this simulation, the tuning voltages $\eta_{1,2,3}$ have been set to the values provided by the SAF for $\Delta\varphi=\pi$ rad. As can be seen, the transient simulation presents discrepancies with the phase shift values predicted by the SAF technique. These discrepancies can be explained by the presence of distributed elements in the circuits. Since the SAF uses a frequency domain description of the array, which is provided by the HB system, this formulation makes use of frequency domain models of all the distributed elements. In nonlinear microwave circuits, these components are more easily described and analyzed in the frequency domain, especially in the case of dispersive transmission lines. Then, when evaluating the steady state of these circuits, the frequency domain description typically provides more reliable simulation results than the time-domain modeling of these elements, which is used by the transient simulation \cite{suarez:libro1,suarez:libro2}.
\

\begin{figure}[!h] 
\fontsize{8}{10}\selectfont
\centering 
 \subfloat[]{\includegraphics[width=3in]{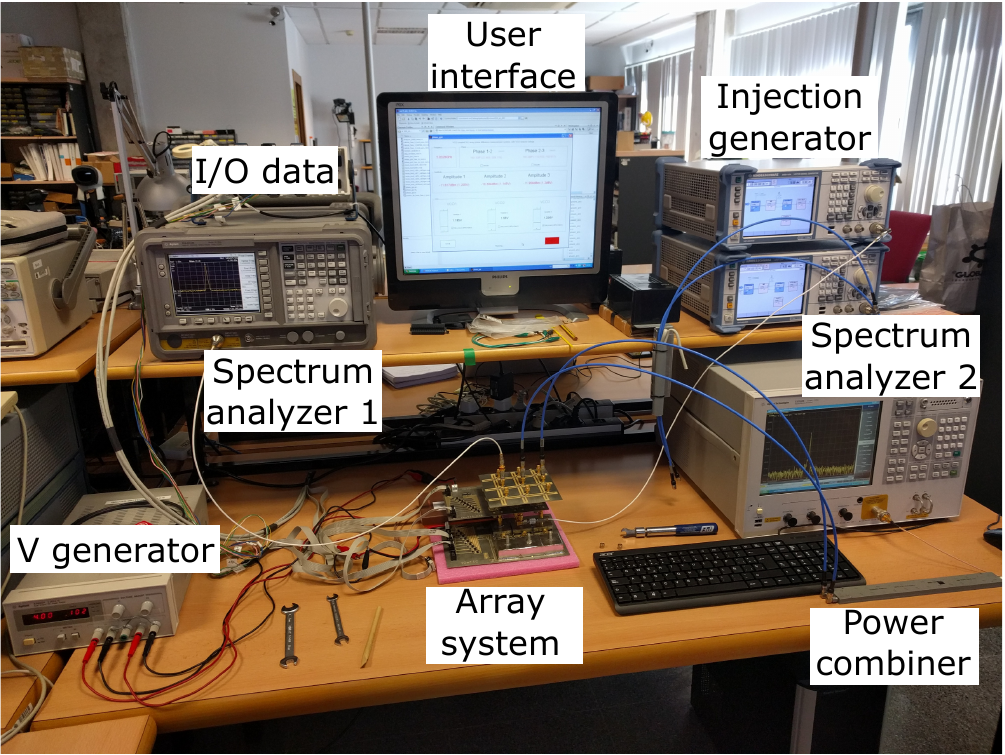}}
\hfill\\
 \subfloat[]{\includegraphics[width=3in]{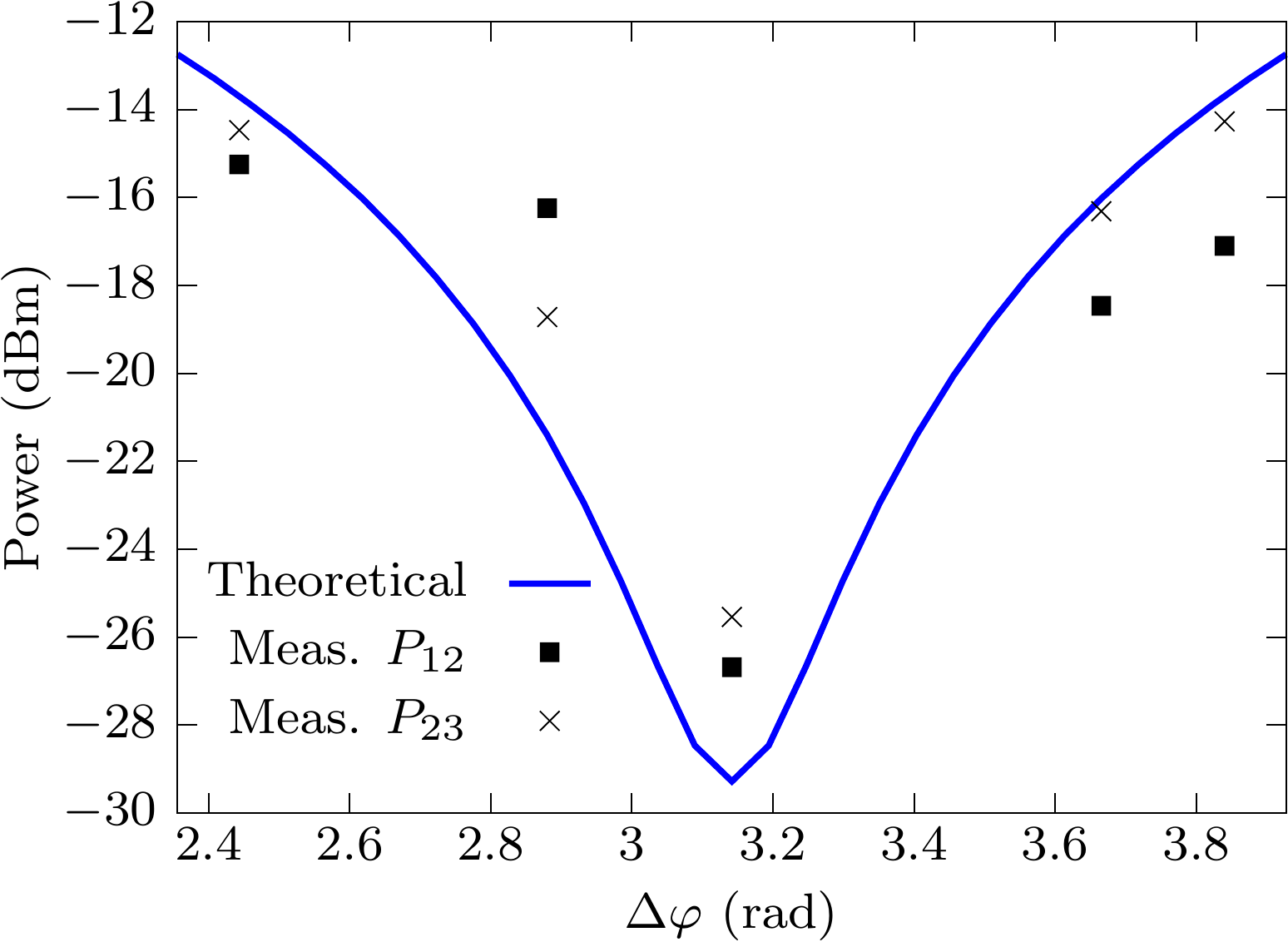}}\\
\hfill\\
\caption{ Free-running array. (a) Measurement setup to analyze the effect of the phase shift between the oscillators in the array  (b) Measurement of the signal power resulting from the combination of the output voltage signals of adjacent VCOs as the phase shift $\Delta\varphi$ is varied in the vicinity of $\Delta\varphi=\pi$ rad.}
\label{f11}
\end{figure}

In order to verify that the phase shift obtained in the measurements corresponds to the range obtained in Fig \ref{f6}(a), an additional measuring procedure has been carried out in the new version of the paper. With the aid of this procedure, the power evolution of the signal resulting from the combination of the output voltage signals $v_i$ of adjacent oscillators has been measured. The measurement setup is shown in Fig. \ref{f11}(a). In the first place, the signals $v_1$ and $v_2$ have been combined and the spectrum of the resulting signal $v_{12}$ has been measured in the analyzer 2. In the analyzer 1, the spectrum of $v_3$ is checked to assure that the array remains synchronized. Then, the measured power $P_{12}$ of the signal $v_{12}$ has been measured as the phase shift $\Delta\varphi$ is varied in the vicinity of $\Delta\varphi=\pi$ rad, with the result of Fig. \ref{f11}(b). The same procedure has been repeated for the signal $v_{23}$, resulting from combining the outputs $v_2$ and $v_3$. The measurement results have been compared in the same figure with the theoretical power of the combined signal. As can be observed, the power of the combined signal presents a minimum for $\Delta\varphi=\pi$ rad, corresponding to the signals out of phase. Note that the signals do not cancel each other completely, since the output powers of adjacent VCOs are not equal. In addition, the attenuators, the wires and the power combiner introduce additional phase shifts whose effect we have tried to minimize.

\subsubsection{Analysis of the injected array}
In the second place, the piecewise technique has been applied to analyze the injected array. For this purpose, an external generator has been injected to the VCO with index $q=2$, modeled by a current source of amplitude $I_s=0.5\ mA$. For a given value of the constant phase shift $\Delta\varphi$, the set of frequency $\omega_s$ values for which the array gets synchronized to the input source will be called the synchronization range. This range can be obtained by solving system (\ref{e5}) as the input phase $\theta_s$ is swept in the range $[0,2\pi]$ rad. 

\begin{figure} [!h] 
\fontsize{8}{10}\selectfont
    \centering
\includegraphics[width=3in]{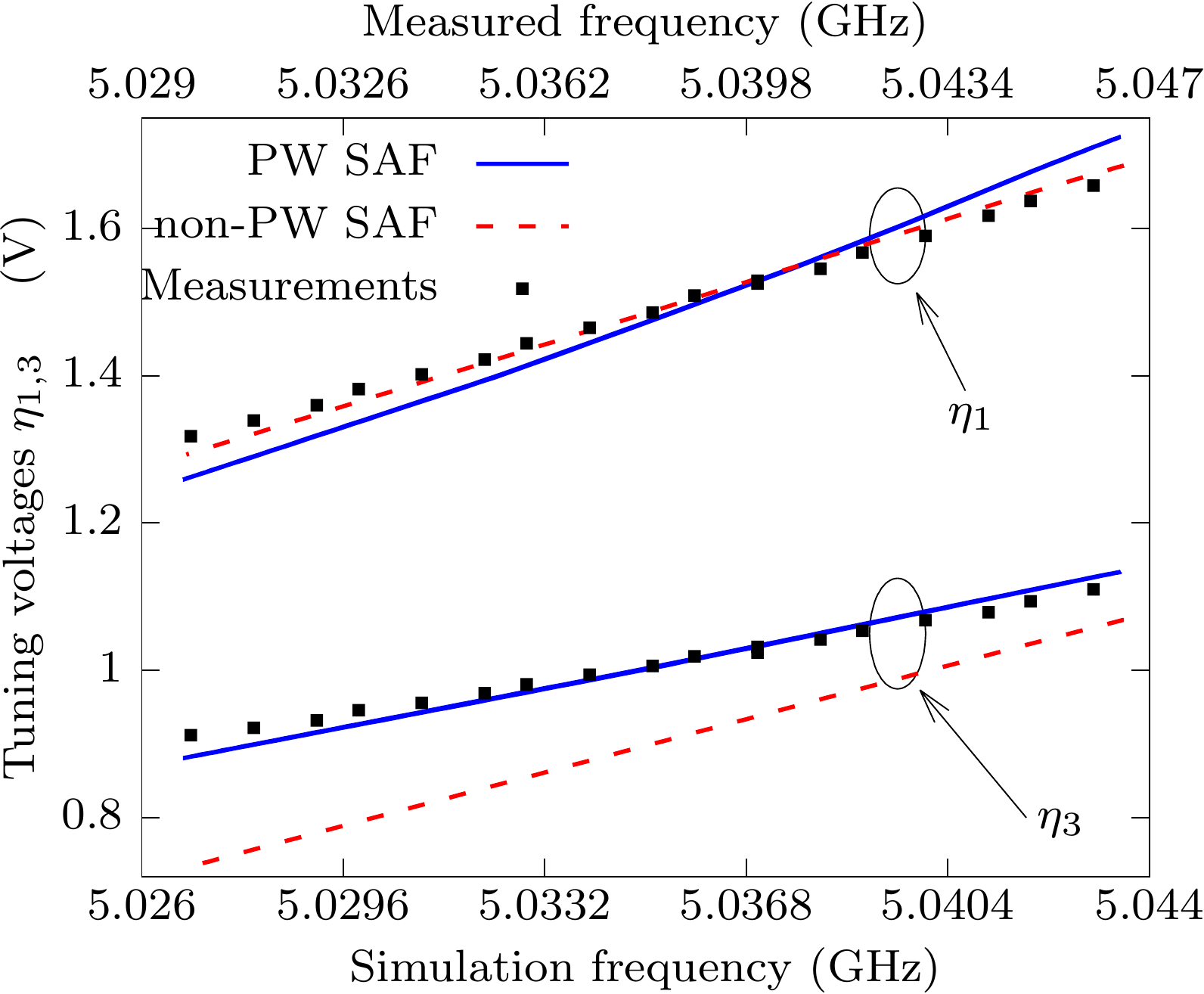}
 \caption{Injected array. Variation of the tuning voltages $\eta_{1,3}$ versus the synchronization frequency $f_s$. The constant phase shift has been fixed to $\Delta\varphi=7\pi/6$ rad and the input generator phase $\theta_s$ has been swept in the range $\C{0,2\pi}$ rad.}
  \label{f7}
\end{figure}

The constant phase shift has been fixed to $\Delta\varphi=7\pi/6$ rad to work within the stable region measured in Fig. \subref*{f6a}. The stability of each synchronized solution has been analyzed in a similar fashion to the previous case. In Fig. \ref{f7}, the tuning voltages $\eta_{1,3}(\theta_s)$ have been represented versus the injection frequency $\omega_s(\theta_s)$. A slight relative shift of $0{.}6\%$ has been observed in the measured frequencies. These curves provide the necessary values of the tuning voltages of the outermost VCOs in the array to maintain a constant phase shift. The results obtained with the non-PW SAF technique have been superimposed. As in the free-running case, the PW SAF shows a better agreement with measurements than the non-PW SAF as the tuning voltage values are farther from the acquiescent point $\eta_2=1{.}55\ V$. 

\section{Conclusion}
A new technique has been presented for the simulation of the synchronized solutions in arrays of coupled oscillators. The technique makes use of piecewise linear admittance functions to model each element of the array. These functions take into account the whole VCO characteristic, performing a global analysis that allows the maintenance of accuracy along the whole synchronization interval. The technique has been validated by comparison with circuit-level HB simulations in a free-running array of Van der Pol-type oscillators. Then, it has been applied to a coupled system of FET based oscillators at 5 GHz, in which the circuit-level HB technique is unable to converge. The synchronization ranges have been successfully predicted both under free-running conditions and injection-locked to an external generator. These results have been validated by comparison with measurements. In all cases, the new piecewise technique has shown a better accuracy than the previous non-piecewise one.

{\appendices
\section{Calculation of the synchronized solution}
In this appendix, the procedure detailed in Section \ref{s5} for the calculation of the synchronized solution providing a constant phase-shift between the oscillators in the array is schematized:

\begin{enumerate}
\item Discretize the tuning range of each oscillator taking $p$ samples $\eta_1^c,\ldots,\eta_p^c$.
\item For each $i-th$ oscillator, calculate the derivatives $I_{iG^{r,i}}^k$ and $Y_{ix}^k$, defined in (\ref{e7}) and (\ref{e2}), where $x=V_i,\omega_i,\eta_i$. These derivatives are calculated by means of an AG, as explained in Section II-A.
\item Build system (\ref{e5}) from (\ref{e4}) with the set of unknowns $\LL{\V{V},\V{\eta},\omega_s}$, where $\eta_q$ is fixed to a constant value. 
\item Solve system (\ref{e5}) for the given constant phase shift value $\Delta\varphi$ through an optimization procedure. At each evaluation of the function $\V{H}$, the intervals $\C{\eta_{k_i}^c,\eta_{k_{i+1}}^c},\ i=1,\ldots,N$ containing the values of the tuning voltages $\eta_i,\ i=1,\ldots,N$ determine the coefficients $k_i$ of the derivatives $I_{iG^{r,i}}^{k_i}$ and $Y_{ix}^{k_i}$.
\end{enumerate}

This simulation procedure has been followed to obtain the synchronized solutions of the coupled oscillator system versus the constant phase-shift in Figs. \ref{f3} and \ref{f6}(b).
}


\begin{IEEEbiography}[{\includegraphics[width=1in,height=1.25in,clip,keepaspectratio]{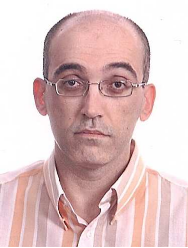}}]{Pedro Umpierrez} 
was born in Las Palmas de Gran Canaria, Spain, in 1979. He received the Telecommunication Engineering degree in 2006 and the Ph.D. degree in the same discipline in 2016, both from the Universidad de Las Palmas de Gran Canaria (ULPGC), Las Palmas, Spain. In 2007, he joined the Signals and Communications Department, ULPGC, where he was involved in projects related with RF circuits simulation and design and IoT communication systems. His research interests include coupled oscillators arrays analysis for beam steering.

\end{IEEEbiography}

\vskip -3\baselineskip plus -1fil

\begin{IEEEbiography}[{\includegraphics[width=1in,height=1.25in,clip,keepaspectratio]{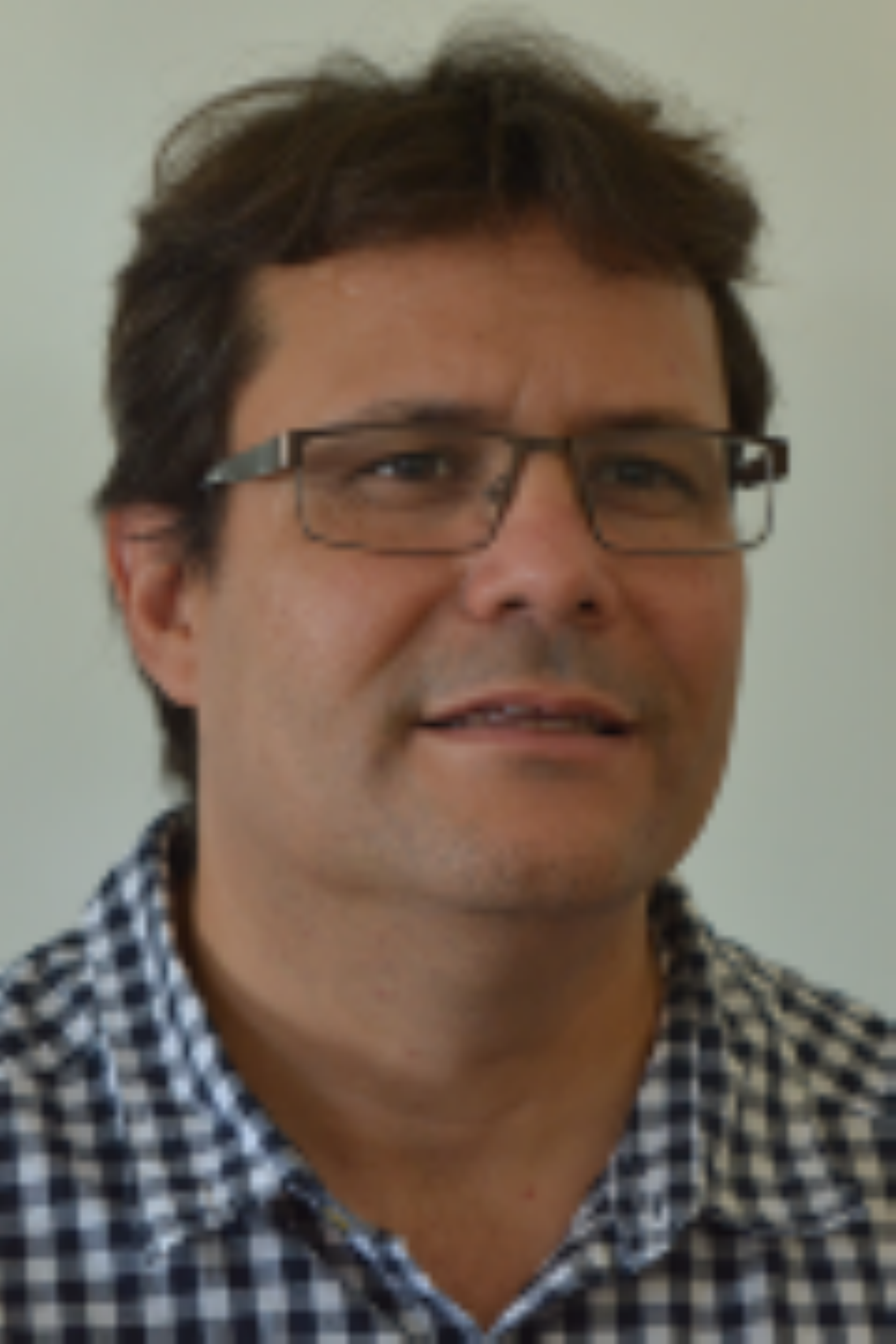}}]{Victor Ara\~na}
(A’94–M’05) was born in Las Palmas de Gran Canaria, Spain, in 1965. He received the Telecommunication Engineering degree from the Universidad Politecnica de Madrid, Madrid, Spain in 1990, and the Ph.D. degree from the Universidad de Las Palmas de Gran Canaria (ULPGC), Las Palmas, Spain, in 2004. He is currently an Assistant Professor with the Signal and Communication Department, ULPGC. He has been the leading Researcher in several
Spanish research and development projects and has taken part in a number of Spanish and European projects in collaboration with industries. His research interests include the nonlinear design of microwave circuits, control subsystem units, and communications systems applied to data acquisition complex networks.

\end{IEEEbiography}

\vskip -3\baselineskip plus -1fil

\begin{IEEEbiography}[{\includegraphics[width=1in,height=1.25in,clip,keepaspectratio]{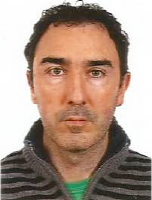}}]{Sergio Sancho}
(A’04–M’04) received the degree in Physics from Basque Country University in 1997. In 1998 he joined the Communications Engineering Department of the University of Cantabria, Spain, where he received the Ph.D. degree in Electronic Engineering in February 2002. At present, he works at the University of Cantabria, as an Associate Professor of its Communications Engineering Department. His research interests include the nonlinear analysis of microwave autonomous circuits and frequency synthesizers, including stochastic and phase-noise analysis.

\end{IEEEbiography}

\ifCLASSOPTIONcaptionsoff
  \newpage
\fi

\IEEEtriggercmd{\enlargethispage{-5in}}


\begin{thebibliography}{10}
\providecommand{\url}[1]{#1}
\csname url@samestyle\endcsname
\providecommand{\newblock}{\relax}
\providecommand{\bibinfo}[2]{#2}
\providecommand{\BIBentrySTDinterwordspacing}{\spaceskip=0pt\relax}
\providecommand{\BIBentryALTinterwordstretchfactor}{4}
\providecommand{\BIBentryALTinterwordspacing}{\spaceskip=\fontdimen2\font plus
\BIBentryALTinterwordstretchfactor\fontdimen3\font minus
  \fontdimen4\font\relax}
\providecommand{\BIBforeignlanguage}[2]{{%
\expandafter\ifx\csname l@#1\endcsname\relax
\typeout{** WARNING: IEEEtran.bst: No hyphenation pattern has been}%
\typeout{** loaded for the language `#1'. Using the pattern for}%
\typeout{** the default language instead.}%
\else
\language=\csname l@#1\endcsname
\fi
#2}}
\providecommand{\BIBdecl}{\relax}
\BIBdecl

\bibitem{Lynch:CO}
J.~J. Lynch and R.~A. York, ``Synchronization of oscillators coupled through
  narrow-band networks,'' \emph{IEEE Trans. Microw. Theory Tech.}, vol.~49,
  no.~2, pp. 237--249, Feb. 2001.

\bibitem{Heath:beam}
T.~Heath, ``Simultaneous beam steering and null formation with coupled,
  nonlinear oscillator arrays,'' \emph{IEEE Trans. Antennas Propag.}, vol.~53,
  no.~6, pp. 2031--2035, 2005.

\bibitem{pogorzelski:continuum}
R.~J. Pogorzelski, ``Continuum modeling of coupled oscillator arrays with
  coupling delay,'' \emph{Radio Science}, vol.~43, no.~04, pp. 1--12, Aug.
  2008.

\bibitem{moussounda:dissertation}
R.~Moussounda, \emph{Analysis and design of coupled-oscillator arrays for
  microwave systems}.\hskip 1em plus 0.5em minus 0.4em\relax Dissertation. The
  Ohio State University, 2013.

\bibitem{pogorzelski:book:coupled}
R.~J. Pogorzelski and A.~Georgiadis, \emph{Coupled-Oscillator Based
  Active-Array Antennas}.\hskip 1em plus 0.5em minus 0.4em\relax Wiley, 2011.

\bibitem{kijsanayotin:coupled}
T.~Kijsanayotin, J.~Li, and J.~F. Buckwalter, ``A 70-{GHz LO} phase-shifting
  bidirectional frontend using linear coupled oscillators,'' \emph{IEEE Trans.
  Microw. Theory Tech.}, vol.~65, no.~3, pp. 892--904, March 2017.

\bibitem{qi_phased_array}
L.~Qi, Q.~Xie, L.~Deng, and Z.~Wang, ``A novel terahertz phased array based on
  coupled oscillators,'' in \emph{2018 IEEE MTT-S International Wireless
  Symposium (IWS)}, May 2018, pp. 1--3.

\bibitem{simeone:wireless}
O.~Simeone, U.~Spagnolini, Y.~Bar-Ness, and S.~H. Strogatz, ``Distributed
  synchronization in wireless networks,'' \emph{IEEE Signal Process. Mag.},
  vol.~25, no.~5, pp. 81--97, Sept 2008.

\bibitem{ponton:wireless:ims}
M.~Pont\'on and A.~Su\'arez, ``Stability analysis of wireless
  coupled-oscillator circuits,'' in \emph{2017 IEEE MTT-S International
  Microwave Symposium (IMS)}, June 2017, pp. 83--86.

\bibitem{ponton:wireless:mtt}
M.~Pont\'on, A.~Herrera, and A.~Su\'arez, ``Wireless-coupled oscillator systems
  with an injection-locking signal,'' \emph{IEEE Trans. Microw. Theory Tech.},
  vol. (Early access), pp. 1--17, 2018.

\bibitem{suarez:OA:noise}
A.~Suarez, F.~Ramirez, and S.~Sancho, ``Stability and noise analysis of
  coupled-oscillator systems,'' \emph{IEEE Trans. Microw. Theory Tech.},
  vol.~59, no.~4, pp. 1032--1046, Apr. 2011.

\bibitem{suarez:OA:general}
A.~Suarez, S.~Sancho, and F.~Ramirez, ``General formulation for the analysis of
  injection-locked coupled-oscillator systems,'' \emph{IEEE Trans. Microw.
  Theory Tech.}, vol.~61, no.~12, pp. 4730--4744, Dec. 2013.

\bibitem{suarez:OA:global}
A.~Su\'arez, F.~Ram\'irez, S.~Sancho, and J.~M. Collantes, ``Global stability
  analysis of coupled-oscillator systems,'' \emph{IEEE Trans. Microw. Theory
  Tech.}, vol.~63, no.~1, pp. 165--180, Jan. 2015.

\bibitem{suarez:libro2}
A.~Suarez, \emph{Analysis and Design of Autonomous Microwave Circuits}.\hskip
  1em plus 0.5em minus 0.4em\relax Wiley, 2009.

\bibitem{ramirez:divider}
F.~Ramirez, E.~de~Cos, and A.~Suarez, ``Nonlinear analysis tools for the
  optimized design of harmonic-injection dividers,'' \emph{IEEE Trans. Microw.
  Theory Tech.}, vol.~51, no.~6, June, 2003.

\bibitem{sancho:cycle_slips}
S.~Sancho, F.~Ram\'irez, and A.~Su\'arez, ``Stochastic analysis of cycle slips
  in injection-locked oscillators and analog frequency dividers,'' \emph{IEEE
  Trans. Microw. Theory Tech.}, vol.~62, no.~12, pp. 3318--3332, Dec. 2014.

\bibitem{ponton:modes}
M.~Pont\'on and A.~Su\'arez, ``Oscillation modes in symmetrical wireless-locked
  systems,'' \emph{IEEE Trans. Microw. Theory Tech.}, vol.~66, no.~5, pp.
  2495--2510, May 2018.

\bibitem{ramirez:chirped:tmtt}
F.~Ramírez, S.~Sancho, M.~Pontón, and A.~Suárez, ``Two-scale envelope-domain
  analysis of injected chirped oscillators,'' \emph{IEEE Trans. Microw. Theory
  Tech.}, vol.~66, no.~12, pp. 5449--5461, Dec 2018.

\bibitem{adler:locking}
R.~Adler, ``A study of locking phenomena in oscillators,'' \emph{Proc. IRE},
  vol.~34, no.~6, pp. 351--357, June 1946.

\bibitem{ramirez:sincro_noise}
F.~Ram\'irez, M.~Pont\'on, S.~Sancho, and A.~Su\'arez, ``Phase-noise analysis
  of injection-locked oscillators and analog frequency dividers,'' \emph{IEEE
  Trans. Microw. Theory Tech.}, vol.~56, no.~2, Feb. 2008.

\bibitem{VCO_Array_7}
H.~C. Chang, X.~Cao, U.~K. Mishra, and R.~A. York, ``Phase noise in coupled
  oscillator theory and experiment,'' \emph{IEEE Trans. Microw. Theory Tech.},
  vol.~45, no.~5, pp. 604--615, May 1997.

\bibitem{VCO_Array_1}
R.~Pogorzelski, ``A 5-by-5 element coupled oscillator-based phased array,''
  \emph{IEEE Trans. Antennas Propag.}, vol.~53, no.~4, pp. 1337--1345, Apr.
  2005.

\bibitem{york:quiasioptical}
R.~A. {York}, ``Nonlinear analysis of phase relationships in quasi-optical
  oscillator arrays,'' \emph{IEEE Transactions on Microwave Theory and
  Techniques}, vol.~41, no.~10, pp. 1799--1809, Oct 1993.
  
\bibitem{kurokawa:free}
K.~Kurokawa, ``Some basic characteristics of broadband negative resistance
  oscillator circuits,'' \emph{The Bell System Technical Journal}, vol.~48,
  no.~6, pp. 1937--1955, July 1969.

\bibitem{arana:measurement}
B.~P\'erez, V.~Arana, J.~P\'erez-Mato, and F.~Cabrera, ``360 degree phase
  detector cell for measurement systems based on switched dual multipliers,''
  \emph{IEEE Trans. Microw. Wireless Compon. Lett.}, vol.~27, no.~5, pp.
  503--505, May 2017.

\bibitem{keysight:oscport}
\BIBentryALTinterwordspacing
``Keysight: Using oscport.'' [Online]. Available:
  \url{http://edadocs.software.keysight.com/pages/viewpage.action?pageId=5919753#HarmonicBalanceforOscillatorSimulation-UsingOscPort}
\BIBentrySTDinterwordspacing

\bibitem{collado:coupled:HB}
A.~Collado, F.~Ramirez, A.~Suarez, and J.~Pascual, ``Harmonic-balance analysis
  and synthesis of coupled-oscillator arrays,'' \emph{IEEE Microw. Wireless
  Compon. Lett}, vol.~14, no.~5, p. 192, May 2004.

\bibitem{suarez:libro1}
A.~Su\'arez and R.~Qu\'er\'e, \emph{Stability Analysis of Nonlinear Microwave
  Circuits}.\hskip 1em plus 0.5em minus 0.4em\relax Boston: Artech-House, 2003.

\end{thebibliography}
\end{document}